\documentclass[pre,reprint,showpacs,floatfix,superscriptaddress]{revtex4-1}
\usepackage{graphicx,bm,amssymb,amsmath,dcolumn}

\usepackage{epstopdf}   
\usepackage{multirow}
\usepackage{bbm}
\usepackage{color}
\usepackage{graphicx}
\usepackage{hyperref}
\usepackage{subfigure}
\usepackage{threeparttable}
\usepackage{color}
\usepackage{amsthm}
\newtheorem{lemma}{Lemma}
\newcommand{\<}{\langle}
\renewcommand{\>}{\rangle}

\newcommand{\var}{\mathrm{var}}
\newcommand{\cov}{\mathrm{cov}}
\newcommand{\bbN}{\mathbb{N}}
\newcommand{\sO}{\mathcal{O}}
\newcommand{\bbP}{\mathbb{P}} 

\newcommand{\bbE}{\mathbb{E}}
\newcommand{\dc}{d_{\mathrm{c}}}
\newcommand{\pc}{p_{\mathrm{c}}}
\newcommand{\yp}{y_{\parallel}}
\newcommand{\sN}{\mathcal{N}}
\newcommand{\fN}{\mathfrak{N}}
\newcommand{\sS}{\mathcal{S}}
\newcommand{\sR}{\mathcal{R}}
\newcommand{\sW}{\mathcal{W}}
\newcommand{\sA}{\mathcal{A}}
\newcommand{\fS}{\mathfrak{S}}
\newcommand{\nupar}{\nu_\parallel}

\newcommand{\be}{\begin{equation}}
\newcommand{\ee}{\end{equation}}

\begin{document}
\title{\texorpdfstring{High-precision Monte Carlo study of directed percolation in $(d+1)$ dimensions}{High-precision Monte Carlo study of directed percolation in (d+1) dimensions}}
\author{Junfeng Wang}
\affiliation{Hefei National Laboratory for Physical Sciences at Microscale and Department of Modern Physics, University of Science and Technology of China, Hefei, Anhui 230026, China}
\affiliation{School of Electronic Science and Applied Physics, Hefei University of Technology, Hefei 230009, China}
\author{Zongzheng Zhou}
\affiliation{School of Mathematical Sciences, Monash University, Clayton, Victoria~3800, Australia}
\author{Qingquan Liu}
\affiliation{Hefei National Laboratory for Physical Sciences at Microscale and Department of Modern Physics, University of Science and Technology of China, Hefei, Anhui 230026, China}
\author{Timothy M. Garoni}
\email{tim.garoni@monash.edu}
\affiliation{School of Mathematical Sciences, Monash University, Clayton, Victoria~3800, Australia}
\author{Youjin Deng}
\email{yjdeng@ustc.edu.cn}
\affiliation{Hefei National Laboratory for Physical Sciences at Microscale and Department of Modern Physics, University of Science and Technology of China, Hefei, Anhui 230026, China}

\date{\today}
\begin{abstract}
 We present a Monte Carlo study of the bond and site directed (oriented) percolation models in $(d+1)$ dimensions on simple-cubic and body-centered-cubic lattices, with $2 \leq d \leq 7$.
 A dimensionless ratio is defined, and an analysis of its finite-size scaling produces improved estimates of percolation thresholds.
 We also report improved estimates for the standard critical exponents.
 In addition, we study the probability distributions of the number of wet sites and radius of gyration, for $1 \leq d \leq 7$.
\end{abstract}

\maketitle

\section{Introduction}
 Directed (or oriented) percolation (DP) is a fundamental model in non-equilibrium statistical mechanics.
 A variety of natural phenomena can be modeled by DP, including forest fires~\cite{BroadbentHarmmersley1957,Albano1994}, epidemic diseases~\cite{Mollison1977}, 
 and transport in porous media~\cite{BouchaudGeorges1990,HavlinBenavraham1987}.

 A major reason for the longstanding interest in DP is its conjectured universality, first described by Janssen~\cite{Janssen1981} and Grassberger~\cite{Grassberger1982}.
 Specifically, it is believed that any model possessing the following properties will belong to the DP universality class: short-range interactions;
 a continuous phase transition into a unique absorbing state; a one-component order parameter and no additional symmetries.

 At and above the upper critical dimension ($\dc=4$), mean-field values for the critical 
 exponents $\beta=1$, $\nu_{\parallel} = 1$, $\nu_{\perp} = 1/2$ are believed to hold.
 For $d<\dc$ however, no exact results for either critical exponents or thresholds are known, and instead one relies on numerical estimates
 obtained by series analysis, transfer matrix methods, and Monte Carlo simulations.
 In $(1+1)$ dimensions, series analysis~\cite{Jensen1996,Jensen1999} has enabled the threshold estimates on several lattices to be determined to the eighth decimal place, with the critical
 exponents being estimated to the sixth decimal place.

 Estimates of thresholds and critical exponents for $d\geq2$ can be found
 in~\cite{GrassbergerZhang1996,Grassberger2009a,PerlsmanHavlin2002,LubeckWillmann2004,AdlerBergerDuarteMeir1988,Blease1977,Grassberger2009b}.
 Compared with results for $d=1$ however, the precision of these estimates in higher dimensions is less satisfactory.
 The central undertaking of the present work is to use high-precision Monte Carlo simulations to
 systematically study the thresholds of bond and site DP on simple-cubic (SC) and body-centered-cubic (BCC) lattices for $2\leq d\leq 7$.
 
 In order to obtain precise estimates of the critical thresholds,
 we study the finite-size scaling of the dimensionless ratio $Q_t = N_{2t}/N_t$, where $N_t$ is the mean number of sites becoming wet at time $t$.

 Having obtained these estimates for $p_c$, we then fix $p$ to our best estimate of $\pc$ and use finite-size scaling to obtain improved estimates of the critical exponents for $d=2,3$.
 In addition, we also study the finite-size scaling at $p_c$ of the distribution
 \begin{equation}
   p_{\sN}(t,s):=\bbP(\sN_t=s|\sN_t>0),
   \label{N distribution}
 \end{equation}
 where $\sN_t$ is the number of sites becoming wet at time $t$.
 We conjecture, and numerically confirm, that
 \begin{equation}
   p_{\sN}(t,s) \sim t^{-y_{\sN}}F_{\sN}(s/t^{y_{\sN}}), \qquad t\to\infty,
   \label{eq:dist_N}
 \end{equation}
 with exponent $y_{\sN} = \theta + \delta$, where $\theta = (d\nu_\perp-\beta)/\nupar$ and $\delta = \beta/\nupar$ for $d <\dc$ and $y_{\sN}=1$ for $d \geq \dc$.
  We also study an analogous distribution of the random radius of gyration, as in Eq.~(\ref{eq:dist_N}) with $y_{\sN}$ 
 being replaced by $y_{\sR}= \nu_{\perp}/\nu_{\parallel}$.

 The remainder of this paper is organized as follows.
 Section~\ref{Description of Simulations} introduces the DP models we study and describes how the simulations were performed.
 Results are presented in Secs.~\ref{Percolation Thresholds}, \ref{Critical Exponents} and~\ref{Critical Distributions}.
 We conclude with a discussion in Sec.~\ref{sec6}.
 In Appendix~\ref{d=1 results} we present estimated thresholds of bond and site DP on the square, triangle,
 honeycomb and kagome lattices, while Appendix~\ref{proofs relating to estimators} contains some technical results justifying the definitions of the improved estimators defined in Sec.~\ref{Improved Estimators}.

\section{Description of the model and simulations}
\label{Description of Simulations}
\subsection{Generating DP configurations}
\label{Generating DP configurations}
 \begin{figure}
   \centering
   \includegraphics[scale=0.30]{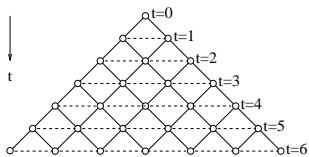}
   \caption{Stochastic formulation of DP on the square lattice. 
   The vertical direction corresponds to time, and the dashed lines identify the sets $V_t$.
   }
   \label{squareLatticeDiagram}
 \end{figure}
Although DP was originally introduced from a stochastic-geometric perspective~\cite{BroadbentHarmmersley1957}, as the natural analog of isotropic percolation to oriented lattices,
the most common formulation of DP is as a stochastic cellular automaton.
To obtain a stochastic formulation of DP on a given oriented lattice, one defines a sequence $(V_t)_{t\ge0}$ which partitions the set of lattice sites, 
such that each adjacent site directed to $v\in V_t$ belongs to some $V_{t'}$ with $t'<t$. 
See Fig.~\ref{squareLatticeDiagram} for the example of the square lattice.
By setting $V_0=\{\mathbf{0}\}$, the trajectory of the stochastic process then generates the cluster connected to the origin.
Typically $t'=t-1$, and the resulting process is then Markovian.

For both site and bond DP, at time $t$ the stochastic process visits each site $v\in V_t$ and 
sets either $s_v=1$ (wet) or $s_v=0$ (dry).
In more detail, the process proceeds as follows.
At $t=0$, we wet the origin with probability 1.
At time $t>0$, we construct for each $v\in V_t$ the (random) set $E_v$ of edges directed from wet sites to $v$.
In the case of site DP, if $E_v$ is non-empty we set $s_v=1$ with probability $p$, otherwise we set $s_v=0$.
For bond DP, we select an edge $e\in E_v$ and occupy it with probability $p$. If $e$ is occupied, we set $s_v=1$, and then proceed to update the next site in $V_t$.
If $e$ is unoccupied, we repeat the procedure for the next edge in $E_v$, and continue until either an edge is occupied or the set $E_v$ is 
exhausted~\footnote{We note that the version of bond DP that we are simulating generates a different ensemble of bond configurations compared to the standard geometric version of bond DP, in which each edge is 
  occupied independently. However, the resulting site configurations generated by these two bond DP models are identical. 
  Since we only consider properties of the site configurations in this article, the distinction is unimportant for our purposes.
  For the sake of computational efficiency, we find the version described in the text more convenient.}.

 We note that in this description, the sets $V_t$ have been given a pre-specified order, as have the sets of edges incident to each $v\in V_t$.
 The precise form of these orderings is obviously unimportant, and in practice they were induced in the natural way from the coordinates of the vertices.
 We used a hash table~\cite{Sedgewick98} to store the wet sites in our simulations, as described in~\cite{Grassberger2003}.

 For $p>\pc$, there is a non-zero probability that the number of wet sites will diverge as $t\to\infty$.
 In our simulations, the cluster growth stops either at the first time that no new sites become wet, or when $t=t_{\rm max}$, where $t_{\rm max}$ is predetermined.
 The values of $t_{\max}$ used for each simulation were chosen as follows. For site and bond DP with $2\le d\le 5$, we set $t_{\max}=2^{14}$.
 On SC lattice with $d=6,7$, we set $t_{\max}=2^{13}$ and $t_{\max}=2^{11}$ respectively. 
 On BCC lattice with $d=6,7$, we set $t_{\max}=2^{12}$ and $t_{\max}=2^{10}$ respectively.
 In all cases, the number of independent samples generated was $10^9$.

\subsection {Lattices}
\label{Lattices}
\begin{figure}
\centering
\subfigure{
\includegraphics[scale=0.40]{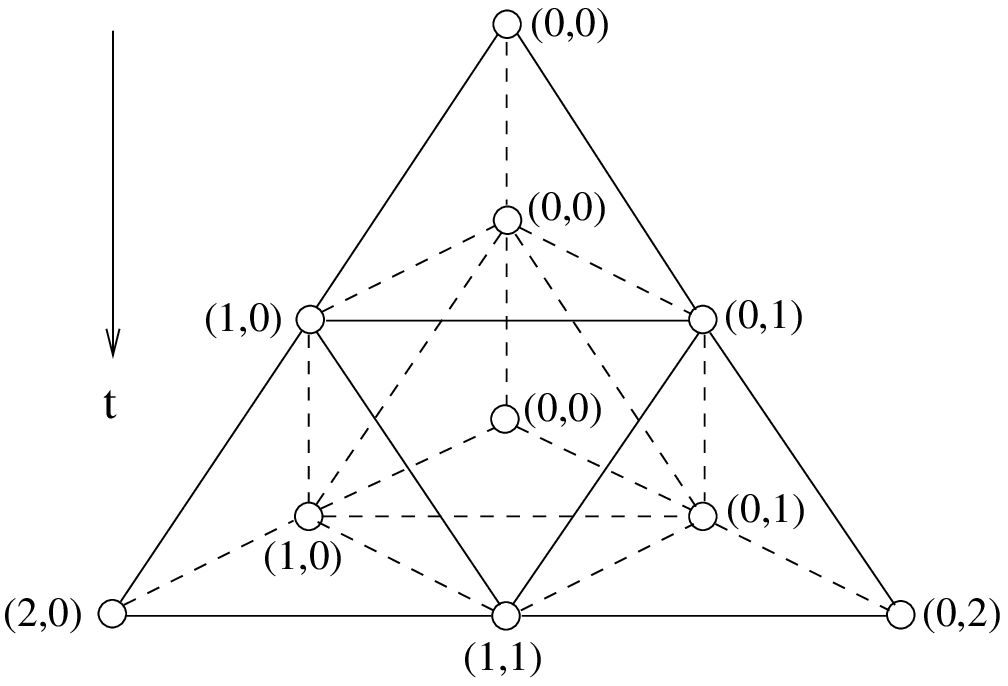}}
\subfigure{
\includegraphics[scale=0.40]{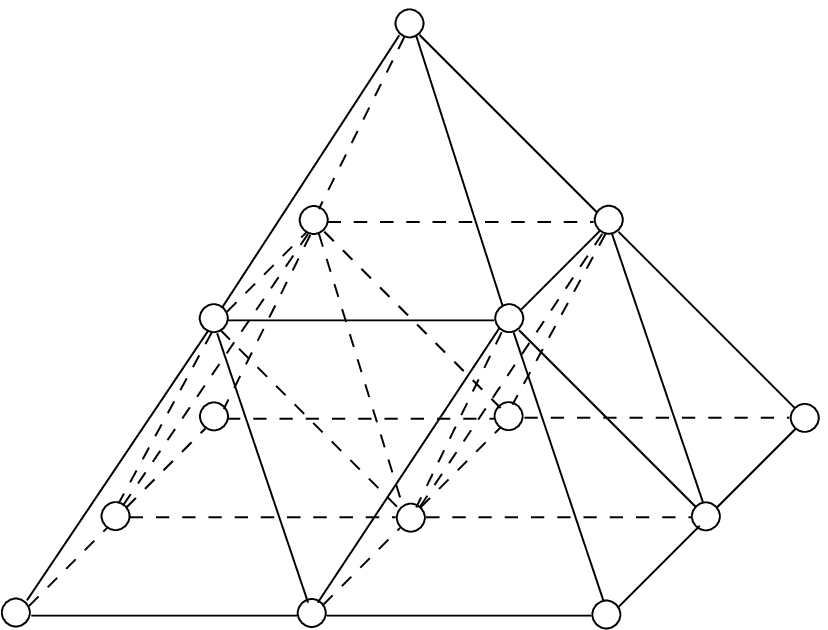}}
\caption{(2+1)-dimensional SC (left) and BCC (right) lattices.}
\label{SCClatticeBCClatticeDiagram}
\end{figure}
We simulated $(d+1)$-dimensional simple-cubic (SC) and body-centered cubic (BCC) lattices with $2\le d \le 7$.
The stochastic processes formulation of DP on these lattices that we used in our simulations is Markovian, 
and is described most easily by explicitly describing the sets $V_t$ together with the edges between $V_t$ and $V_{t+1}$.
In $(d+1)$ dimensions, each $V_t\subset\mathbb{Z}^d$. Let $\mathbf{x}\in V_t$, and let $\{\mathbf{e}_1,\ldots,\mathbf{e}_d\}$ denote the standard basis of $\mathbb{Z}^d$.
On the BCC lattice, the coordinates of the $\lambda=2^d$ neighbors of $\mathbf{x}$ in $V_{t+1}$ are $\mathbf{x}+\sum_{i=1}^d\alpha_i\,\mathbf{e}_i$ for $\boldsymbol{\alpha}\in\{0,1\}^d$.
On the SC lattice, the coordinates of the $\lambda=d+1$ neighbors of $\mathbf{x}$ in $V_{t+1}$ are $\mathbf{x}+\sum_{i=1}^d\alpha_i\,\mathbf{e}_i$ for all $\boldsymbol{\alpha}\in\{0,1\}^d$ with 
$\|\boldsymbol{\alpha}\|_1\le1$.
The (2+1)-dimensional cases are illustrated in Fig.~\ref{SCClatticeBCClatticeDiagram}.

 \subsection{Observables}
 \label{Observables}
 For each simulation we sampled the following random variables:
 \begin{enumerate}
 \item $\sN_t$, the number of sites becoming wet at time $t$;
 \item $\sS_t = \sqrt{\sum_{v} r_v^2}$, where $r_v$ denotes the Euclidean distance of the site $v$ 
  to the time axis,  and the sum is over all wet sites in $V_t$;
 \item $\fN_t = \sum_{v\in V_{t}} b_v$, where $b_v$ is the number of Bernoulli trials 
  needed to determine the state of $v\in V_{t}$, given the configuration of sites in $V_{t-1}$;
 \item $\fS_t = \sqrt{\sum_{v\in V_{t}} b_v\,r_v^2}$.
 \end{enumerate}
 We note that, as shown in Appendix~\ref{proofs relating to estimators}, we have
 \begin{align}
   \left\< {\sN_{t}}\right\> &= p\left\< {\fN_{t}}\right\>,\label{N alternative estimator identity}\\
   \left\< {\sS_{t}^2}\right\> &= p\left\< {\fS_{t}^2}\right\>,\label{R alternative estimator identity}
 \end{align}
 where $\< \cdot \>$ denotes the ensemble average. As explained in Section~\ref{Improved Estimators}, $\fN_t$ and $\fS_t$ can be used to construct reduced-variance estimators.
 
 Using the above random variables, we then estimated the following quantities:
 \begin{enumerate}
\item The percolation probability $P_t = \bbP(\sN_t>0)$;
\item The mean number of sites becoming wet at time $t$, $N_t = \< \sN_t \>$;
\item The dimensionless ratio $Q_t = N_{2t}/N_t$;
\item The radius of gyration $R_t^2 = \left\< \sS_t^2\right\>/N_t$;
\item The distribution $p_{\sN}(t,s)$ defined by \eqref{N distribution};
\item The distribution
  \begin{equation}
    p_{\sR}(t,s) := \bbP(\sR_t = s\,|\,\sN_t >0)
    \label{R distribution}
  \end{equation}
  where
  \begin{equation}
    \sR_t := 
    \begin{cases}
    \displaystyle\frac{\sS_t}{\sqrt{\sN_t}}, & \sN_t>0,\\
    0, & \sN_t = 0.
    \end{cases}
  \end{equation}
\end{enumerate}
We expect the second moment of $p_{\sR}(t,\cdot)$ to display the same critical scaling as the radius of gyration.
We discuss this point further in Sec.~\ref{Critical Distributions}.

 \subsection{Improved Estimators}
 \label{Improved Estimators}
 To estimate $N_t$, $R_t^2$ and $Q_t$, we adopted the variance reduction technique introduced in~\cite{Grassberger2003,Grassberger2009b,FosterGrassbergerPaczuski2009}, the details of which we now describe.
 To clearly distinguish sample means generated by our simulated data from the ensemble averages to which they converge, 
 we will use $\overline{X}=\sum_{i=1}^n X^{(i)}/n$ to denote the sample mean of $n$ independent realizations $X^{(1)},\ldots,X^{(n)}$ of the random variable $X$.
 While $\lim_{n\to\infty}\overline{X}=\< X\>$, we emphasize that $\overline{X}$ is a random variable for any finite $n$.

 In addition to the naive estimator $\overline{\sN}_t$, we can also estimate $N_t$ via
 \be
 \widehat{N_t} := p^t\prod_{t'=1}^{t} \frac{\overline{\fN}_{t'}} {\overline{\sN}_{t'-1}}.
 \label{eq:improvedN}
 \ee
 Indeed, taking the number of samples to infinity and using~\eqref{N alternative estimator identity} we find
 \begin{equation}
   \widehat{N_t} = p^t\prod_{t'=1}^{t} \frac{\overline{\fN}_{t'}} {\overline{\sN}_{t'-1}}
   \longrightarrow
  \prod_{t'=1}^{t} \frac{N_{t'}}{N_{t'-1}} = N_t.
 \end{equation}

 Any convex combination of $\overline{\sN}_t$ and the estimator~\eqref{eq:improvedN} will therefore also be an estimator for $N_t$. As our final estimator for $N_t$ we therefore used
\begin{equation}
\alpha \overline{\sN}_t +(1-\alpha) \widehat{N_t},
\label{final N estimator}
\end{equation}
with $\alpha=\alpha_{\min}$ chosen so as to minimize the variance of~\eqref{final N estimator}. Explicitly,
  \be
  \alpha_{\min} = \frac{\var(\widehat{N_t}) - \cov(\overline{\sN}_t,\widehat{N_t})}{\var(\overline{\sN}_t) + \var(\widehat{N_t}) - 2\cov(\overline{\sN}_t,\widehat{N_t})}.
  \label{eq:alpha}
  \ee
  Note that $\alpha_{\min}$ can be readily estimated from the simulation data.
  Similarly, to estimate $Q_t$ we use the minimum-variance convex combination of $\overline{\sN}_{2t}/\overline{\sN}_t$ and $\widehat{N_{2t}}/\widehat{N_t}$.

 An analogous estimator for $R_t^2$ can also be constructed:
 \be
 \widehat{R_t^2} = \sum_{t'=1}^{t}\left(\frac{\overline{\fS}_{t'}^2}{\overline{\fN}_{t'}} - \frac{\overline{\sS}_{t'-1}^2}{\overline{\sN}_{t'-1}}\right)\;,
 \ee
 Taking the number of samples to infinity and using~\eqref{R alternative estimator identity} shows 
 that indeed $\widehat{R_t^2}\to R_t^2$. 
 Analogously to the argument above, we then take the convex combination of $\widehat{R_t^2}$ and 
 $\overline{\sS^2_t}/\overline{N_t}$ with minimum variance to be our final estimator for $R_t^2$.

 We now comment on the motivation behind these definitions.
 For DP on a $\lambda$-ary tree we have the simple identity $\fN_t = \lambda \sN_{t-1}$, 
 which implies that $\widehat{N}_t$ is deterministic in this case, and therefore has precisely zero variance.
 For DP on a $(d+1)$-dimensional lattice, as $d$ increases the updates become more and more like the updates for DP on the $\lambda$-ary tree, and so intuitively one expects 
 that the variance of $\widehat{N}_t$ should decrease as $d$ increases. This is indeed what we observe numerically.
 For the simulations of bond DP on the BCC lattice for example, we find that for $d=4$ the variance of $\widehat{\sN_t}$ is $\approx 0.1$ of the variance of $\overline{\sN}_t$.
 This factor reduces to $10^{-4}$ for $d=7$. 
 For low dimensions, however, the above variance reduction technique is less effective.
 Similar arguments and observations apply to the reduced-variance estimator for the radius of gyration.
 Interestingly, our data suggest that the above technique is more effective for bond DP than site DP.

 \begin{figure}[htb]
 \centering
 \includegraphics[scale=0.60]{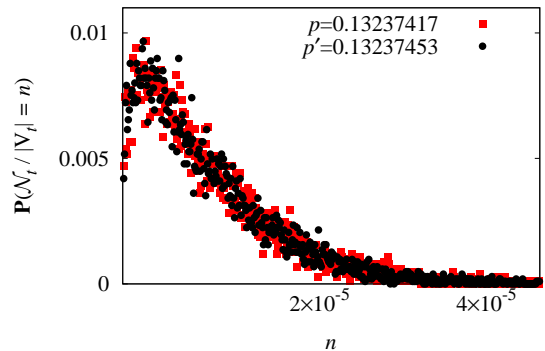}
 \caption{(Color online) Plot of $\bbP(\sN_t/|V_t|=\cdot)$ at $t=16384$ for $d=3$ bond DP on BCC 
  lattice, with $p=0.132\,374\,17$ (square) and $p'=0.132\,374\,53$ (circle).}
 \label{pn}
 \end{figure}

 \begin{figure}[htb]
 \centering
 \includegraphics[scale=0.60, trim = 0 0 0 0, clip]{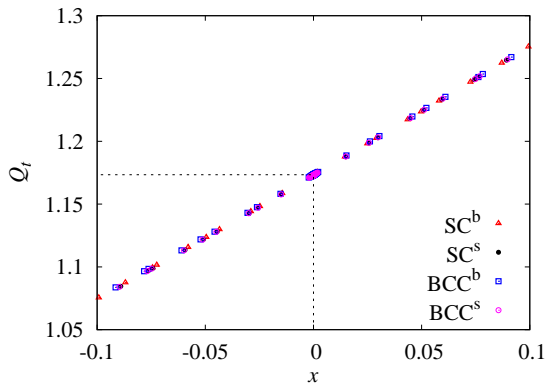}
 \caption{(Color online) Plot of the $Q_t$ data for bond and site DP on the SC and BCC lattices versus $x=q_1(\pc-p)t^{\yp}$ for $d=2$.
 }
 \label{Q_u}
 \end{figure}

\begin{table*}[htbp]
\begin{center}
\begin{tabular}[t]{l|lllllll|l}
\hline
         & $\pc$             & $Q_c$       & $\yp$             & $q_1$        & $q_2$       &  $c_1$   &  $y_u$     & $t_{\rm {min}}$/DF/$\chi^2$ \\
\hline
 $\rm{SC}^b_2$  &0.382\,224\,62(2) & 1.173\,42(4) & 0.776\,7(4)  & $-3.127(7)$  & $2.25(3)$   & $-0.011(1)$ & $-0.5$     &64/224/175 \\
 $\rm{SC}^s_2$  &0.435\,314\,10(5) & 1.173\,42(2) & 0.777\,4(3)  & $-2.403(5)$  & $1.37(2)$   & $-0.025(6)$ & $-0.48(7)$ &64/223/168 \\
 $\rm{BCC}^b_2$ &0.287\,338\,37(2) & 1.173\,36(2) & 0.776\,2(4)  & $-4.177(9)$  & $4.26(4)$   & $-6(3)$   & $-2.1(2)$    &64/224/235 \\
 $\rm{BCC}^s_2$ &0.344\,574\,01(4) & 1.173\,41(2) & 0.777\,2(3)  & $-2.879(6)$  & $1.95(2)$   & $-0.31(2)$& $-0.54(7)$   &64/223/82 \\
\hline
 $\rm{SC}^b_3$  &0.268\,356\,28(1) & 1.076\,52(8) & 0.905(4)     & $-3.8(2)  $  & $5.6(3) $   & $ ~~0.005(1)$&$-0.3    $   &64/223/166 \\
%-0.2(3)
 $\rm{SC}^s_3$  &0.303\,395\,39(2) & 1.075\,2(4)  & 0.906(4)     & $-2.7(1)  $  & $2.9(2) $   & $ ~~0.024(1)$&$-0.26(2)$   &64/223/367 \\
%-0.5(2)
 $\rm{BCC}^b_3$ &0.132\,374\,169(3)& 1.076\,29(6) & 0.904(2)     & $-8.5(2)  $  & $28.6(9)$   & $-0.032(3)$&$-0.62(3)$   &64/359/341 \\
%1.2(9)
 $\rm{BCC}^s_3$ &0.160\,961\,28(1) & 1.076\,7(3)  & 0.904(4)     & $-5.0(2)$    & $10.0(6)$   & $ ~~0.026(1)$&$-0.34(2)$   &64/223/163 \\
%-0.8(4)
\hline
\end{tabular}
\end{center}
\caption{
Fit results for $Q_t$ with $d=2,3$ on the SC and BCC lattices. Superscripts $b$ and $s$ represent bond and site DP, respectively. The subscript represents the dimensionality $d$.}
\label{tablefitqn23d}
\end{table*}

\begin{table}[htbp]
\begin{center}
\begin{tabular}[t]{c|llll|l}
\hline
             & $p_c$        & $q_1$      & $c_1$       & $h_1$      & $t_{\rm {min}}$ \\
\hline
$\rm{SC}^b_4$  & 0.207\,918\,153(3)      & $-5.2(9)$     & $-0.22(2) $ & ~~0.50(2)    & 64   \\
$\rm{SC}^s_4$  & 0.231\,046\,861(3)      & $-3.83(1)$    & $-0.53(2)$  & $-1.2(1)$    & 64   \\
$\rm{BCC}^b_4$ & 0.063\,763\,395(1)      & $-18.6(1)$    & $-0.08(1)$  & ~~3.34(3)    & 64    \\
$\rm{BCC}^s_4$ & 0.075\,585\,154(2)      & $-11.4(4)$    & $-0.55(2)$  & $-1.12(2)$   & 64   \\
\hline
\end{tabular}
\end{center}
\caption{Fit results for $Q_t$ with $d=4$ on the SC and BCC lattices. Superscripts $b$ and $s$ represent bond and site DP, respectively. The subscript represents the dimensionality $d$.}
\label{tablefitqn4d_log}
\end{table}

\begin{table*}[htbp]
\begin{center}
\begin{tabular}[t]{c|llllll|l}
\hline
               & $\pc$                  & $q_1$         & $q_2$     & $c$       &  $c_1$       & $y_u$       & $t_{\rm {min}}$/DF/$\chi^2$ \\
\hline
 $\rm{SC}^b_5$   & 0.170\,615\,153(1)     & $-5.253(5)$   & 13.5(2)   & $-0.72(7)$  & $0.026(1)$   & $-0.49(1)$  & 48/258/208   \\
 $\rm{SC}^s_5$   & 0.186\,513\,581(2)     & $-4.115(6)$   & 8.6(1)    & $-1.17(9)$  & $0.054(1)$   & $-0.49(1)$  & 48/258/172   \\
 $\rm{BCC}^b_5$  & 0.031\,456\,631\,6(1) & $-30.78(4)$   & 450(6)     & $-1.6(7)$   & $0.009(1)$   & $-0.48(1)$  & 48/248/176   \\
 $\rm{BCC}^s_5$  & 0.035\,972\,542\,1(5)  & $-21.17(5)$   & 164(7)    & $-5.3(7)$   & $0.049(1)$   & $-0.48(1)$  & 48/242/119   \\
\hline
 $\rm{SC}^b_6$   & 0.145\,089\,946\,5(4)  & $-6.538(2) $  & 21.4(2)   &  ~~~~-        & $0.028(1)$   & $-0.99(1)$  & 48/235/147   \\
 $\rm{SC}^s_6$   & 0.156\,547\,177(3)     & $-5.428(4) $  & 13.1(4)   &  ~~~~-        & $0.051(3)$   & $-0.87(2)$  & 64/193/55    \\
 $\rm{BCC}^b_6$  & 0.015\,659\,382\,96(3) & $-63.394(8)$  & 1945(30)  &  ~~~~-        & $0.003(1)$   & $-0.99(3)$  & 48/193/110   \\
 $\rm{BCC}^s_6$  & 0.017\,333\,051\,7(4)  & $-49.33(4) $  & 1343(27)  &  ~~~~-        & $0.043(1)$   & $-0.88(2)$  & 48/198/79    \\
\hline
 $\rm{SC}^b_7$   & 0.126\,387\,509\,0(6)  & $-7.663(2) $  & 28.9(4)   &  ~~~~-        & $0.015(2) $  & $-1.32(3)$  & 32/225/196   \\
 $\rm{SC}^s_7$   & 0.135\,004\,173(2)     & $-6.566(4) $  & 20.8(5)   &  ~~~~-        & $0.092(6) $  & $-1.45(2)$  & 32/225/212   \\
 $\rm{BCC}^b_7$  & 0.007\,818\,371\,82(1) & $-127.63(1)$  & 7557(157) &  ~~~~-        & $0.0007(2)$  & $-1.31(8)$  & 32/176/171   \\
 $\rm{BCC}^s_7$  & 0.008\,432\,989\,5(3)  & $-107.0(2) $  & 3882(1000)&  ~~~~-        & $0.036(5) $  & $-1.29(4)$  & 32/181/84    \\
\hline
\end{tabular}
\end{center}
\caption{
Fit results for $Q_t$ with $d=5,6,7$ on the SC and BCC lattices. Superscripts $b$ and $s$ represent bond and site DP, respectively. The subscript represents the dimensionality $d$.}
\label{tablefitqn567d}
\end{table*}

\section{Percolation Thresholds}
\label{Percolation Thresholds}
\subsection{Fitting Methodology}
To estimate the critical threshold $\pc$ we applied an iterative approach. 
We ran preliminary simulations at several values of $p$ and relatively small values of $t_{\max}$, and used these data to estimate $\pc$ by studying the finite-size scaling of $Q_t$.
Further simulations were then performed at and near the value of $\pc$ estimated in the initial runs, using somewhat larger values of $t_{\max}$.
For both site and bond DP, and for each choice of lattice and dimension, 
this procedure was iterated a number of times before we performed our final high-precision runs at the single value of $p$ which corresponded to the best estimate of $\pc$ obtained in the preliminary 
simulations. For these final simulations we used the values of $t_{\max}$ reported in Section~\ref{Generating DP configurations}.

For computational efficiency, we then used re-weighting to obtain expectations corresponding to multiple values of $p$, from each of our final high-precision runs.
Our approach to re-weighting is similar to that described for the contact process in~\cite{Dickman1999}, 
and relies on the simple observation that for any observable $\sA_t$ we have the identity $\<\sA_t\>_{p'} = \left\<\sW_{p,p'}\,\sA_t \right\>_p$, 
where the random variable $\sW_{p,p'}$ is defined on the space of site configurations $C$ by
$$
\sW_{p,p'}(C) = \frac{\bbP_{p'}(C)}{\bbP_{p}(C)} = \prod_{t=1}^{t_{\max}} (\dfrac{p'}{p})^{\sN_t(C)}(\dfrac{1-p'}{1-p})^{\fN_t(C)-\sN_t(C)}.
$$

As with any application of re-weighting, in practice one must of course be careful that the distributions $\bbP_p(\cdot)$ and $\bbP_{p'}(\cdot)$ have sufficient overlap, so that 
a finite simulation with parameter $p$ will generate sufficiently many samples in the neighbourhood of the peak of $\bbP_{p'}(\cdot)$.
As $t$ increases, the range of acceptable $p'$ values is expected to decrease.
To verify that we had sufficient overlap, for both bond and site DP and for each choice of lattice and dimension, 
we performed additional low-statistics simulations ($10^7$ independent samples, rather than $10^9$)
for the $p'$ values furthest from $p$, and compared the histograms generated at $t=t_{\max}$ for simulations at $p'$ with those generated at $p$. In all cases the overlap was excellent.
Figure~\ref{pn} gives a typical example, showing the estimated distribution $\bbP(\sN_t/|V_t|=\cdot)$ at $t=16384$ for $d=3$ bond DP on the BCC lattice, with $p=0.132\,374\,17$ and $p'=0.132\,374\,53$.

These final high-precision data sets were then used to perform our final fits for $\pc$, which we report 
in Tables~\ref{tablefitqn23d},~\ref{tablefitqn4d_log} and~\ref{tablefitqn567d}.
Specifically, we performed least-squares fits of the $Q_t$ data to an appropriate finite-size scaling ansatz.
As a precaution against correction-to-scaling terms that we failed to include in the chosen ansatz, we imposed a lower cutoff
$t>t_{\rm min}$ on the data points admitted in the fit, and we systematically studied the effect on the $\chi^2$ value of increasing $t_{\rm min}$.
In general, our preferred fit for any given ansatz corresponds to the smallest $t_{\min}$ for which the goodness of fit is reasonable 
and for which subsequent increases in $t_{\min}$ do not cause the $\chi^2$ value to drop by vastly more than one unit per degree of freedom.
In practice, by ``reasonable'' we mean that $\chi^2/\mathrm{DF}\lessapprox 1$, where DF is the number of degrees of freedom.

 In Table~\ref{tablefitqn23d},~\ref{tablefitqn4d_log} and~\ref{tablefitqn567d}, we list the results for our preferred fits for $Q_t$, with $d$ from 2 to 7.
 The superscripts ``b" and ``s" are used in these tables to distinguish the bond and site DP, and the subscript denotes the dimensionality $d$.
 The error bars reported in Tables~\ref{tablefitqn23d},~\ref{tablefitqn4d_log} and~\ref{tablefitqn567d} correspond to statistical error only.
 To estimate the systematic error in our estimates of $\pc$ we studied the robustness of the fits to variations in the terms retained in the fitting ansatz and in $t_{\min}$.
 This produced the final estimates of the critical thresholds shown in Table~\ref{tablefinalpchd}.

\subsection{\texorpdfstring{Results for $d=2,3$}{Results for d=2,3}}
\label{Results for d=2,3}
 Near the critical point $\pc$, we expect that
 \begin{equation}
  Q_t(p)=\tilde{Q}(v t^{\yp}, u t^{y_u}) \;,
 \label{qnscaling}
 \end{equation}
 where $v$ and $u$ represent the amplitudes of the relevant and the leading irrelevant scaling fields, respectively, and $\yp =1/ \nupar$ and $y_u <0$ are the associated renormalization exponents.
 Linearizing $v \approx a_1(\pc - p)$ around $p=p_c$ we can expand $Q_t$ as
 \begin{eqnarray}
  Q_t = Q_c &+& \sum_{k\ge1}q_k{(\pc - p)^kt^{k\yp}} + c(\pc - p)t^{\yp + y_u}
  \nonumber \\
  &+& c_1t^{y_u} +\cdots
 \label{qsepfitqn}
 \end{eqnarray}
 where $Q_c = 2^\theta$ and $q_k = a_1^k \,\frac{\partial^k \tilde{Q}}{\partial v^k}|_{v=0}$.
 It follows that $q_k/(q_1)^k$ is a universal quantity.
 In practice, we neglected terms higher than cubic in the finite-size scaling variable $(\pc-p)t^{\yp}$.

 We fitted our data for $Q_t$ to the ansatz~\eqref{qsepfitqn} as described above, and the results are reported in Table~\ref{tablefitqn23d}.
 From the fits for site DP, we observe that on both the SC and BCC lattices, the leading correction exponent $y_u \approx-0.5$ for $d=2$, and $y_u\approx-0.3$ for $d=3$.
 However, for bond DP on the BCC lattice, the fits yield $y_u\approx -2$ for $d=2$, and $y_u\approx-0.6$ for $d=3$.
 This suggests that, within the resolution of our simulations, the amplitude $c_1$ is consistent with zero in this case.
 For the fits for bond DP on the SC lattice, we could not obtain numerically stable fits with $y_u$ left free, and so we
 instead report the results using correction terms $c_1t^{-0.5} + c_2t^{-2}$ for $d=2$ and $c_1t^{-0.3} + c_2t^{-2}$ for $d=3$.

 For $d=2$, we estimate $Q_c = 1.173\,40(6)$, $\nu_{\parallel} = 1/\yp = 1.287(2)$, and $q_2/{q_1}^2=0.24(1)$.
 For $d=3$, we estimate $Q_c = 1.076(1)$, $\nu_{\parallel} = 1/\yp = 1.104(6)$, and $q_2/{q_1}^2 = 0.40(1)$.

 In Fig.~\ref{Q_u} we plot the $Q_t$ data versus $q_1(\pc-p)t^{\yp}$, for bond and site DP on the two-dimensional SC and BCC lattices.
 We use the estimated value $\yp\approx 0.777$, and $q_1$ and $\pc$ are taken respectively from Table~\ref{tablefitqn23d} and Table~\ref{tablefinalpchd}.
 An excellent collapse is observed in Fig.~\ref{Q_u}. The data for $t <1024$ have been excluded to suppress the effects of finite-size corrections.
 The data collapse to a line with slope 1 clearly demonstrates universality.

%\begin{table*}[htbp]
%\begin{table}
%\begin{center}
%\begin{tabular}[t]{c|llll|l}
%\begin{tabular}{c|llll|l}
%\hline
%             & $p_c$        & $q_1$      & $c_1$       & $h_1$      & $t_{\rm {min}}$/DF/$\chi^2$ \\
%\hline
%$\rm{SC}^b_4$  & 0.207\,918\,153(3)      & $-5.2(9)$     & $-0.22(2) $ & ~~0.50(2)    & 64/130/55    \\
%$\rm{SC}^s_4$  & 0.231\,046\,861(3)      & $-3.83(1)$    & $-0.53(2)$  & $-1.2(1)$    & 64/225/253   \\
%$\rm{BCC}^b_4$ & 0.063\,763\,395(1)      & $-18.6(1)$    & $-0.08(1)$  & ~~3.34(3)    & 64/141/83    \\
%$\rm{BCC}^s_4$ & 0.075\,585\,154(2)      & $-11.4(4)$    & $-0.55(2)$  & $-1.12(2)$   & 64/130/160   \\
%\hline
%\end{tabular}
%\end{center}
%\caption{Fit results for $Q_t$ with $d=4$ on the SC and BCC lattices. Superscripts $b$ and $s$ represent bond and site DP, respectively. The subscript represents the dimensionality $d$.}
%\label{tablefitqn4d_log}
%\end{table}
%\end{table*}

\begin{table*}[htbp]
\begin{tabular}[t]{l|ll|lllll}
\hline
Lattice            & Site               &                                     &   & Bond               &                   \\
                   & $p_c$(Present)     & $p_c$(Previous)                     &   & $p_c$(Present)     & $p_c$(Previous)   \\
\hline
$d=2$, SC         & 0.435\,314\,11(10) & 0.435\,31(7)~\cite{GrassbergerZhang1996}  &   & 0.382\,224\,62(6)  & 0.382\,223(7)~\cite{GrassbergerZhang1996} \\
$d=2$, BCC        & 0.344\,574\,0(2)   & 0.344\,573\,6(3)~\cite{Grassberger2009a}  &   & 0.287\,338\,38(4)  & 0.287\,338\,3(1)~\cite{PerlsmanHavlin2002} \\
                   &                    & 0.344\,575(15)~\cite{LubeckWillmann2004}         &  &                        & 0.287\,338(3)~\cite{GrassbergerZhang1996} \\
\hline
$d=3$, SC         & 0.303\,395\,38(5)   & 0.302\,5(10)~\cite{AdlerBergerDuarteMeir1988}     &   & 0.268\,356\,28(5)  & 0.268\,2(2)~\cite{Blease1977} \\
$d=3$, BCC        & 0.160\,961\,28(3)  & 0.160\,950(30)~\cite{LubeckWillmann2004}  &   & 0.132\,374\,17(2)   &  ~~~~-       \\
\hline
$d=4$, SC         & 0.231\,046\,86(3)  &    ~~~~-                             &   & 0.207\,918\,16(2)  & 0.208\,5(2)~\cite{Blease1977} \\
$d=4$, BCC        & 0.075\,585\,15(1)  & 0.075\,585\,0(3)~\cite{Grassberger2009b}  &   & 0.063\,763\,395(5)  &  ~~~~-      \\
                   &                    & 0.075\,582(17)~\cite{LubeckWillmann2004}         &   &                   &          \\
\hline
$d=5$, SC         & 0.186\,513\,58(2)  &    ~~~~-    &   & 0.170\,615\,155(5)  & 0.171\,4(1)~\cite{Blease1977}\\
$d=5$, BCC        & 0.035\,972\,540(3) & 0.035\,967(23)~\cite{LubeckWillmann2004} &   & 0.031\,456\,631\,8(5)  &  ~~~~-     \\
\hline
$d=6$, SC         & 0.156\,547\,18(1)  &  ~~~~-      &   & 0.145\,089\,946(3)  & 0.145\,8~\cite{Blease1977}\\
$d=6$, BCC        & 0.017\,333\,051(2) &  ~~~~-    &   & 0.015\,659\,382\,96(10)  & ~~~~-    \\
\hline
$d=7$, SC         & 0.135\,004\,176(10) & ~~~~-     &   & 0.126\,387\,509(3) & 0.127\,0(1)~\cite{Blease1977}\\
$d=7$, BCC        & 0.008\,432\,989(2)   & ~~~~-     &   & 0.007\,818\,371\,82(6)  & ~~~~-     \\
\hline
\end{tabular}
\caption{Final estimates of critical thresholds for bond and site DP on the SC and BCC lattices, with $2\leq d \leq 7$.
A dash ``-'' implies that we are unaware of any previous estimates in the literature.}
\label{tablefinalpchd}
\end{table*}

%{\it Result for $d=4$.}
\subsection{\texorpdfstring{Results for $d=4$}{Results for d=4}}
\label{Results for d=4}
 At the upper critical dimension, the existence of dangerous irrelevant scaling fields typically leads to both multiplicative and additive logarithmic corrections to the mean-field behavior.
 Field-theoretic arguments~\cite{JanssenTauber2005,JanssenStenull2004} predict that in the neighborhood of criticality
 \begin{equation}
  N_t \sim \left(\ln \frac{t}{t_0}\right)^{\alpha}\Phi\left((\pc-p)t^{\yp}\left(\ln \frac{t}{t_2}\right)^{-\alpha},\, u\,t^{y_u}\right),
 \label{N_logarithmic2}
 \end{equation}
 with $\alpha=1/6$, $\yp=1$ and $\Phi$ a universal scaling function.
 From~\eqref{N_logarithmic2} we then obtain
 \begin{eqnarray}
  Q_t & = & \left(1+\frac{\ln 2}{\ln t + h_1} \right)^{1/6}+ c (\pc-p) \frac{t^{1+y_u}}{(\ln t+h_2)^{1/6}}  \nonumber \\
  &+& \sum_{k\ge 1} q_k(\pc-p)^k \frac{t^{k}}{(\ln t + h_2)^{ k/6}} + c_1 t^{y_u} + \ldots
 \label{sepfitqn_log}
 \end{eqnarray}
 
 We fitted the $d=4$ data for $Q_t$ to the ansatz~\eqref{sepfitqn_log}, and the results of our preferred fits are reported in Table~\ref{tablefitqn4d_log}.
 In the reported fits, we fixed $c=0$ and $h_2=0$ since performing fits with them left free produced estimates for both which were consistent with zero.
 We could not obtain stable fits with $y_u$ left free, and so the reported fits use $y_u=-1$; the resulting estimate of $\pc$ was robust against variations in the fixed value of $y_u$.
 All $q_i$ with $i\ge3$ were set identically to zero. 
 In addition, to suppress the effects of various higher-order corrections associated with the deviation $|\pc-p|$,
 we only fitted the $Q_t$ data corresponding to $p$ values which were sufficiently close to $\pc$ that $q_2$ was consistent with zero.
 Thus, in Table~\ref{tablefitqn4d_log}, we do not report estimates for $q_2$.

\subsection{\texorpdfstring{Result for $d=5$, $6$, $7$}{Result for d=5,6,7}}
\label{Results for d=5,6,7}
For $d>\dc$, we fitted the data for $Q_t$ to the ansatz~\eqref{qsepfitqn} with $Q_c$ and $\yp$ fixed at their mean-field values~\cite{Hinrichsen2000a}, $Q_c=1=\yp$.
The results are reported in Table~\ref{tablefitqn567d}.
Repeating the fits with $Q_c$ and $\yp$ left free produced estimates in perfect agreement with the predicted values.
For $d=6$ and $7$, leaving the amplitude $c$ free produced estimates consistent with zero, and we therefore omitted this term in the reported fits.

From Table~\ref{tablefitqn567d}, we observe that the universal amplitude $q_2/q_1^2\approx 0.5$ holds for all models in $d=5$, $6$ and $7$ dimensions.
We also observe that the leading correction exponents $y_u$ are $\approx -1/2$, $-1$, $-3/2$ for $d=5$, $6$, and $7$, respectively,
in agreement with the field-theoretic prediction~\cite{JanssenTauber2005} of $y_u=2-d/2$.

\subsection {Summary of thresholds}
\label{Summary of thresholds}
 We summarize our final estimates of the critical thresholds for $2\leq d \leq 7$ in Table~\ref{tablefinalpchd}.
 The error bars in these final estimates of $\pc$ are obtained by estimating the systematic error from a comparison of the results from a number of different fits, 
 varying both the terms retained in the fitting ansatz and the value of $t_{\min}$ used.
 For comparison, we also present several previous estimates from the literature.

\begin{figure*}[htbp]
\centering
\subfigure[ ]{
\includegraphics[scale=0.42]{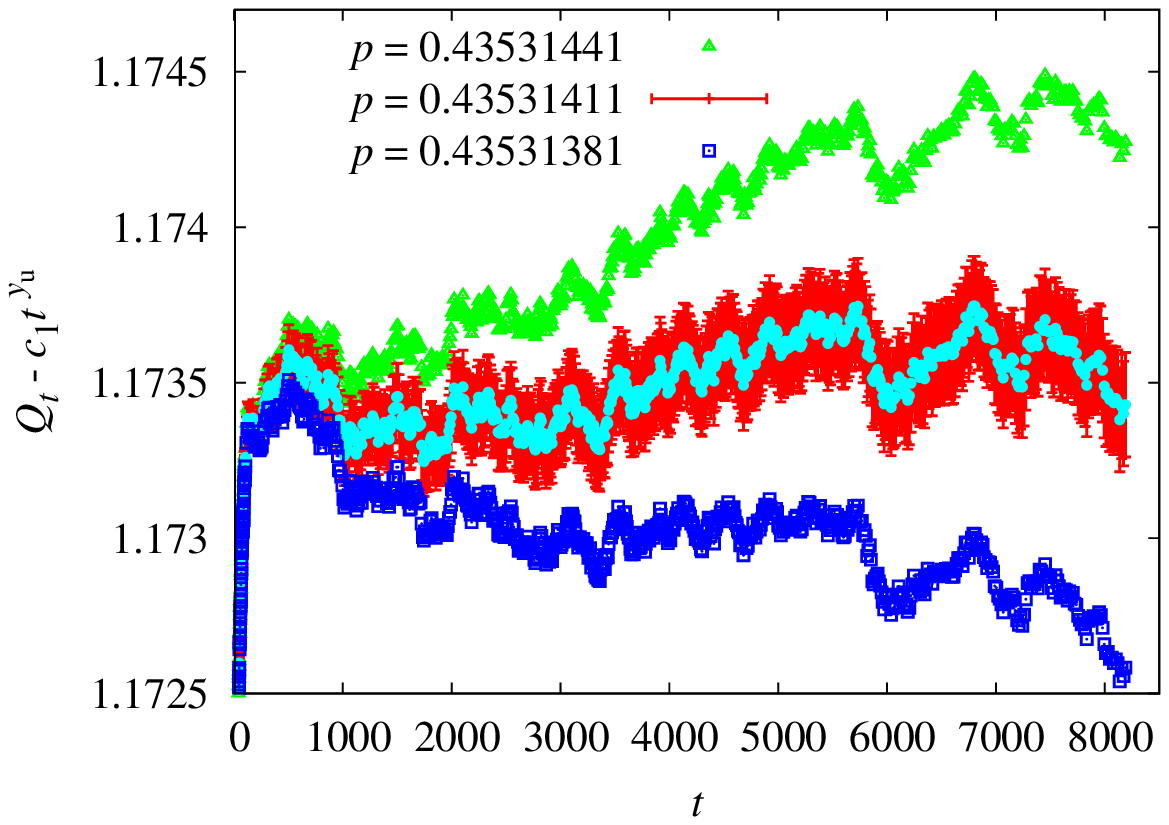}}
\subfigure[ ]{
\includegraphics[scale=0.42]{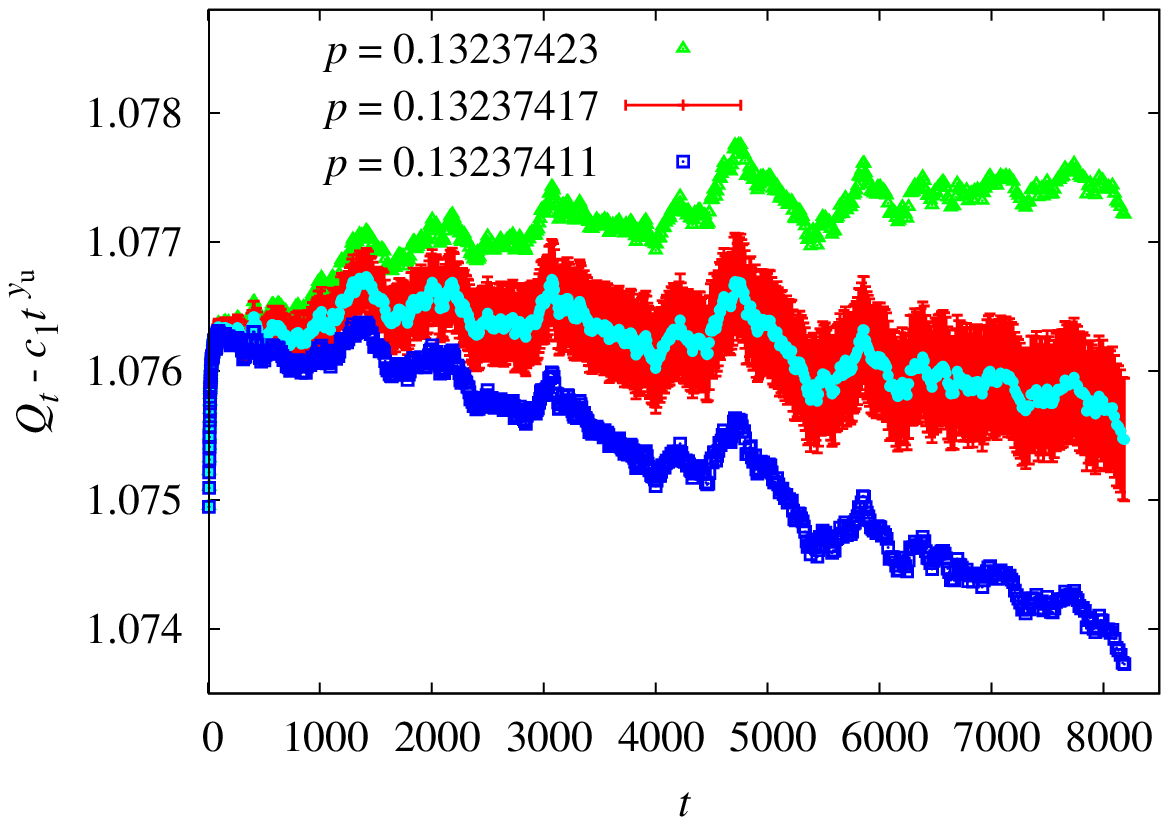}}
\subfigure[]{
\includegraphics[scale=0.42]{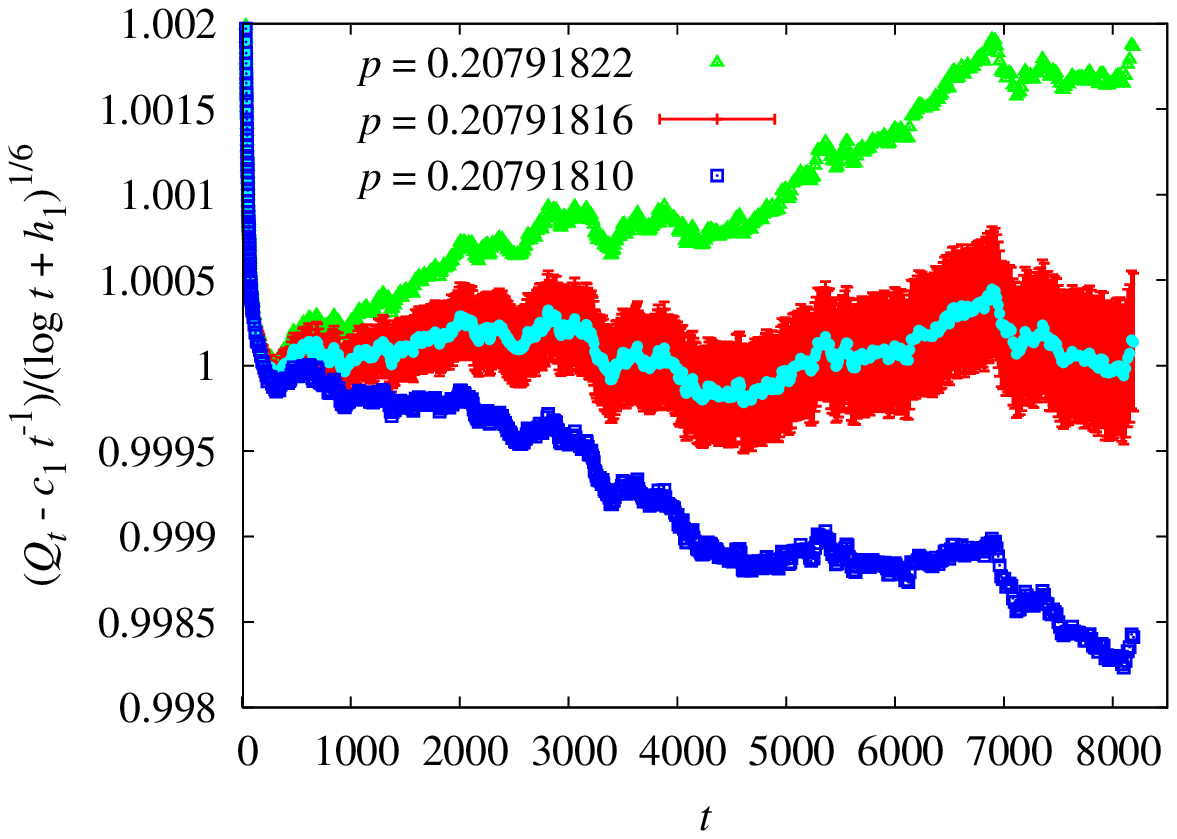}}
\subfigure[ ]{
\includegraphics[scale=0.42]{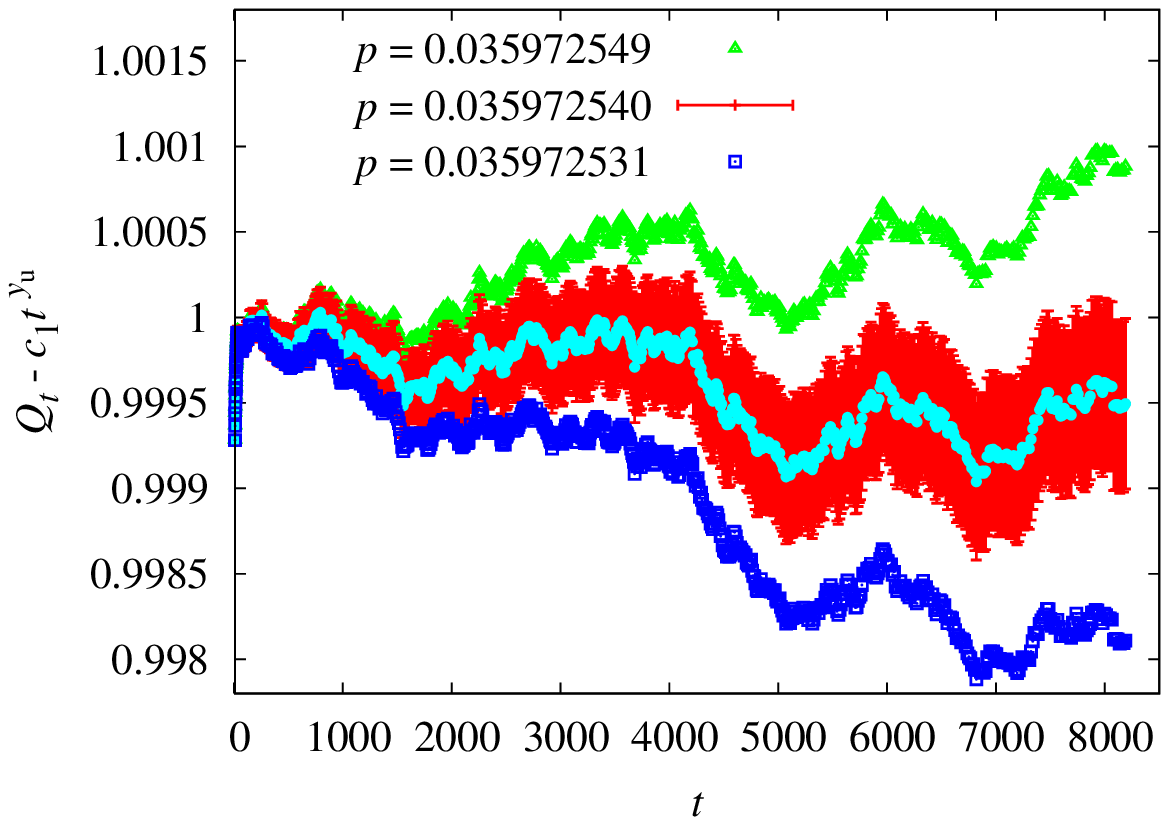}}
\subfigure[ ]{
\includegraphics[scale=0.42]{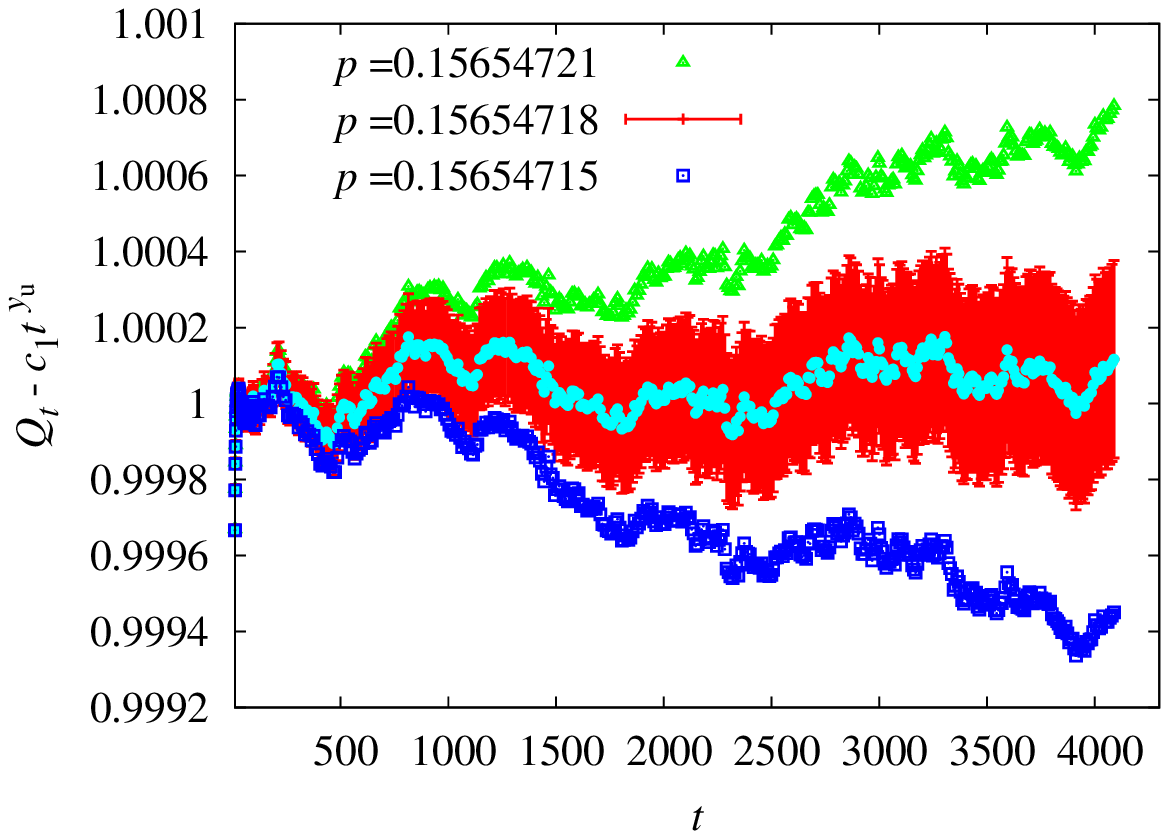}}
\subfigure[ ]{
\includegraphics[scale=0.42]{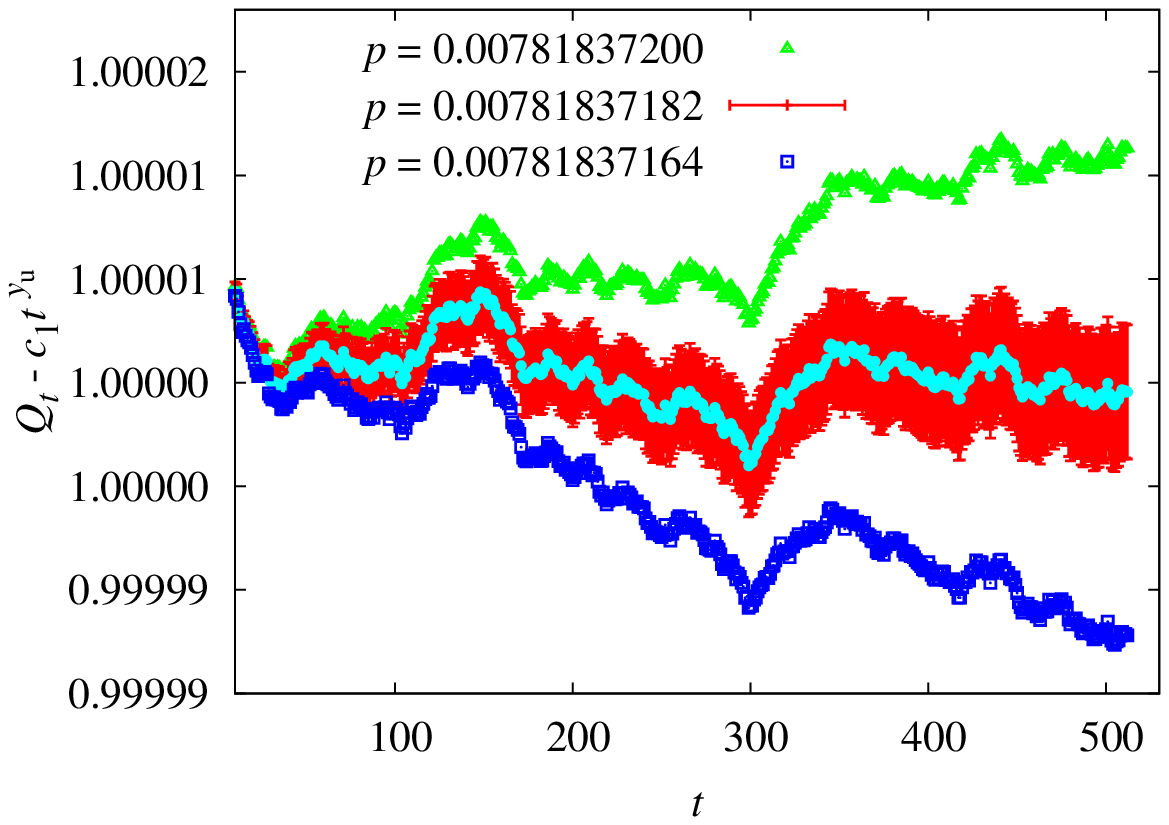}}
\caption{(Color online) Plots of $Q_t-c_1t^{y_u}$ (for $d\neq 4$) and $(Q_t-c_1t^{-1})/(\log t+h_1)^{1/6}$ (for $d=4$) versus $t$ for several DP models.
 The subfigures (a) to (f) respectively correspond to $d=2$ SC site DP, $d=3$ BCC bond DP, $d=4$ SC bond DP, $d=5$ BCC site DP, $d=6$ SC site DP and $d=7$ BCC bond DP.
 The values of $c_1$, $y_u$ and $h_1$ are our best estimates, taken from Tables~\ref{tablefitqn23d},~\ref{tablefitqn4d_log} and~\ref{tablefitqn567d}. 
 The three curves show the Monte Carlo data corresponding to the central value of our estimated $\pc$, and the central value of $\pc$ plus or minus three error bars (from Table~\ref{tablefinalpchd}).
 The curve corresponding to $\pc$ is plotted with its statistical error, corresponding to one standard error.
% The three dashed lines respectively correspond to the central value of our estimated $\pc$, and the central value of $\pc$ plus or minus two error bars (from Table~\ref{tablefinalpchd}).
}
\label{pc_nd}
\end{figure*}

To illustrate the accuracy of our threshold estimates, we plot in Fig.~\ref{pc_nd} the data for $Q_t$ versus $t$ for a number of DP models.
At the critical point, the data for $Q_t$ should tend to a horizontal line as $t$ increases, while the data with $p\neq \pc$ will bend upwards or downwards.
In each case in Fig.~\ref{pc_nd}, the central curve corresponds to our estimated $\pc$,
and the other two curves correspond to the $p$ values which are the estimated $\pc$ plus or minus three error bars.

\begin{figure}[htb]
\centering
\includegraphics[scale=0.45]{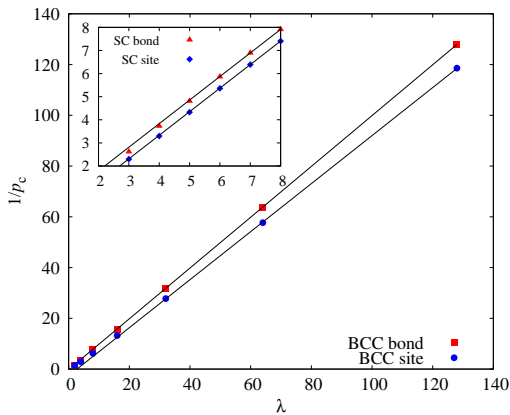}
\caption{(Color online) Plot of $1/\pc$ versus coordination number $\lambda$ for bond and site DP on the BCC lattice.
The lines are obtained by fitting~\eqref{fit_pc} to the $d\geq 4$ data.
The inset shows the analogous plot for the SC lattice.}
\label{figpcz}
\end{figure}

 We conclude this section with some observations regarding the $\pc$ values reported in Table~\ref{tablefinalpchd}.
 Based on empirical observations,~\cite{KurrerSchulten1993} conjectured the ansatz
\begin{equation}
  1/\pc \approx a_1+a_2 \lambda \;,\;\;\mbox{for} \;\;\;\;\lambda \gg 1\;,
  \label{fit_pc}
\end{equation}
relating $\pc$ to the coordination number $\lambda$, when $\lambda$ is large.
In Fig.~\ref{figpcz}, we plot $1/\pc$ versus $\lambda$.
We observe that on the SC lattice, the slopes for bond and site DP are approximately equal, while on the BCC lattice the bond and site cases clearly differ.
In Table~\ref{table_pc_z} we report the values of $a_1$ and $a_2$ obtained by fitting~\eqref{fit_pc} to the $d\geq 4$ data for $\pc$ from Table~\ref{tablefinalpchd}.
From Table~\ref{table_pc_z} we conjecture that $a_2$ is identical for bond and site DP on the SC lattice.

\begin{table}[htbp]
\begin{center}
\begin{tabular}[t]{l|rrrr}
\hline
          & $\rm{SC}^b$  & $\rm{SC}^s$ & $\rm{BCC}^b$  &$\rm{BCC}^s$ \\
\hline
$a_1$     &  $-0.35(4)$     &  $-0.71(2)$   &  $-0.23(4)$       & $-2.6(4)$   \\
$a_2$     &  1.034(5)       &  1.026(2)     &  1.0011(4)        & 0.946(5)   \\
\hline
\end{tabular}
\end{center}
\caption{Estimates of $a_1$ and $a_2$ in~(\ref{fit_pc}), calculated from the $d\ge4$ data.}
\label{table_pc_z}
\end{table}

\begin{table}[htbp]
\begin{center}
\begin{tabular}[t]{l|llll|l}
\hline
$\sO(t)$    & $y_\sO$            & $c_0$         & $c_1$         & $c_2$     & $t_{\rm {min}}$ \\
\hline
%{\multirow{3}{*}{$2$}} &
   $N_t$   & ~~0.230\,70(7)     & 0.976\,0(5)   & 0.004(4)      & ~~4(2)      & 64 \\
   $P_t$   & $-0.451\,1(2)$   & 0.830\,6(8)   & 0.83(9)       & $-30(5)$  & 96 \\
   $R_t^2$ & $~~1.132\,19(4)$   & 1.633\,7(5)   & 1.09(4)       & $-4(2)$   & 64 \\
\hline
   $N_t$   & ~~0.105\,58(10)    & 0.958\,2(7)   & 0.33(5)       & $-4(2)$   & 64  \\
   $P_t$   & $-0.740\,3(3)$   & 1.069(3)      & 0.6(3)        & $-60(12)$ & 64 \\
   $R_t^2$ & $~~1.053\,01(7)$   & 2.715(2)      & 2.2(2)        & $-60(15)$ & 128 \\
\hline
\end{tabular}
\end{center}
\caption{Fits results of $N_t$, $P_t$, and $R_t^2$ on the BCC lattice for $d=2$ (top) 
and 3 (bottom). The leading correction exponent $y_u$ was fixed to $-1$.}
\label{tablefitNPR23d}
\end{table}

\section{Critical Exponents}
\label{Critical Exponents}
 At $p=\pc$, one expects
 \begin{equation}
 P_t\sim t^{-\delta}\;,  \hspace{5mm} N_t\sim t^{\theta}\;, \hspace{5mm} R_t^2\sim t^{2/z} \;.
 \label{eq:scaling}
 \end{equation}
 The critical exponents $\delta$, $\theta$, $z$ are related to the standard exponents $\beta$, $\nu_{\parallel}$, $\nu_{\perp}$ by~\cite{Hinrichsen2000a}
 \begin{equation}
 \label{eq:scaling_relation2}
 \delta = \beta/\nupar \; ,\hspace{2mm} \theta = (d\nu_\perp-2\beta)/\nupar \; ,\hspace{2mm} \mbox{and} \hspace{2mm} z=\nupar/\nu_\perp \; . 
 \end{equation}

Fixing $p$ to our best estimate of $\pc$ from Table~\ref{tablefinalpchd}, we estimated the critical exponents $\theta$, $\delta$, and $z$ for $d = 2$ and $3$,
by studying the critical scaling of $N_t$, $P_t$ and $R_t^2$.
Specifically, we fitted the data for $N_t$, $P_t$, and $R_t^2$ to the ansatz
\begin{equation}
  \sO(t)=t^{y_\sO}(c_0+c_1t^{y_u}+c_2t^{-2}) \;,
  \label{scalingatpc}
\end{equation}
where $y_\sO$ corresponds to $\theta$, $-\delta$ and $2/z$, respectively.
We focused on the case of bond DP on the BCC lattice, since we find empirically that it suffers from the weakest corrections to scaling.
In Table~\ref{tablefitNPR23d}, we report the results of the fits with $y_u$ fixed at $-1$.
To estimate the systematic error in our exponent estimates we studied the robustness of the fits to variations in the fixed value of $y_u$, and in $t_{\min}$.
This produced the final exponent estimates reported in Table~\ref{tablestandardexponentsD23}.

For comparison, we also report in Table~\ref{tablestandardexponentsD23} several previous exponent estimates from the literature.
We note that our estimates of $z$ and $\theta$ in (3+1) dimensions are inconsistent with the field-theoretic predictions reported in~\cite{Janssen1981,BronzanDash1974}.

\begin{figure*}[htbp]
\centering
\subfigure[ ]{\includegraphics[scale=0.42]{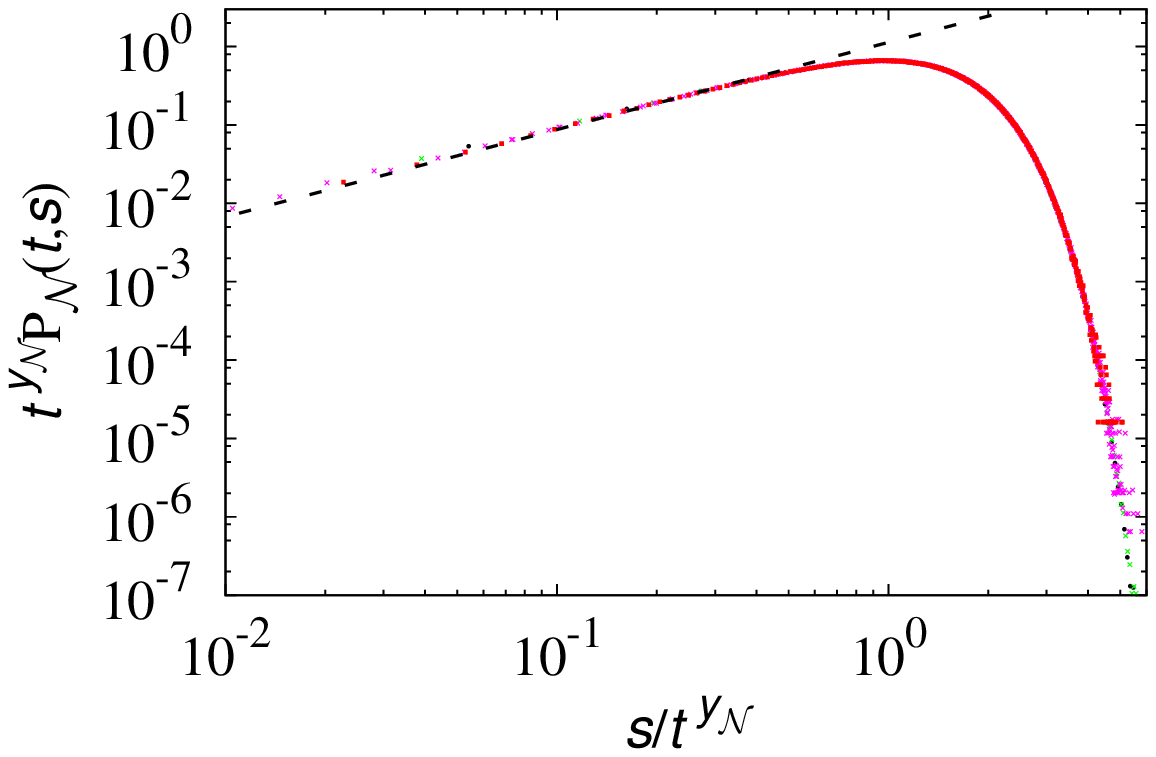}}
\subfigure[ ]{\includegraphics[scale=0.42]{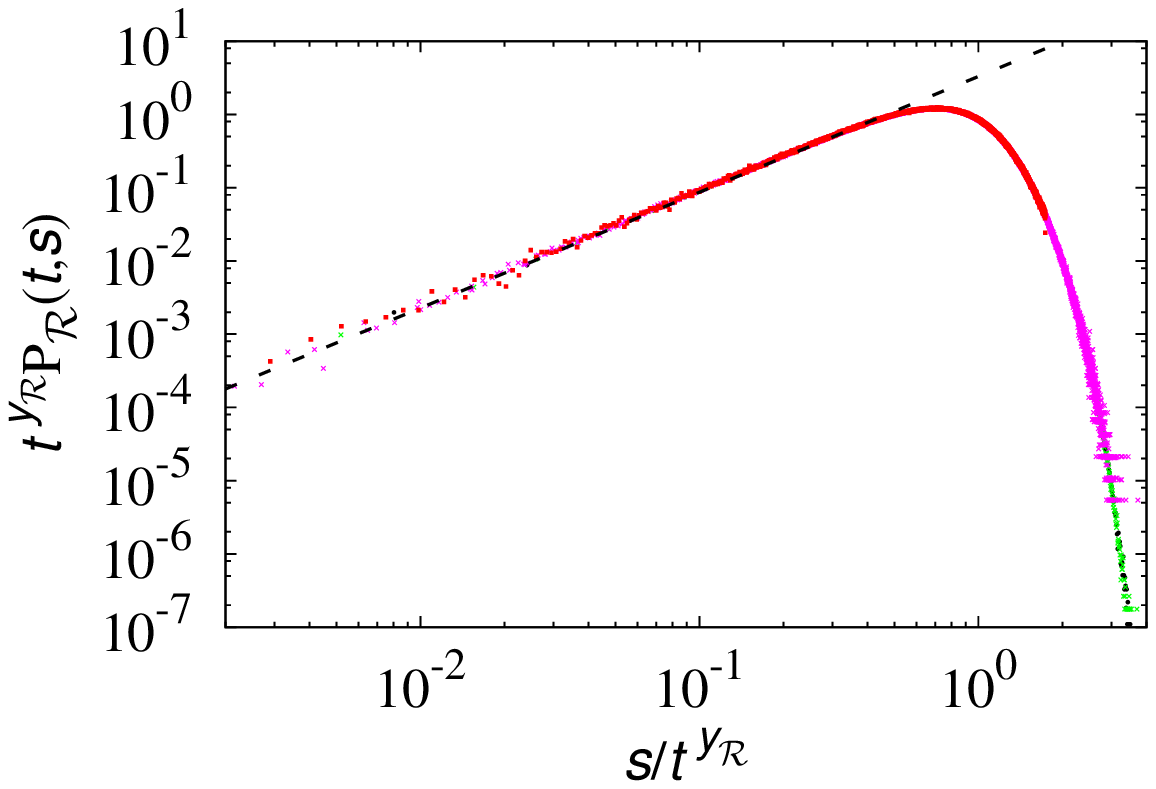}}
\subfigure[ ]{\includegraphics[scale=0.42]{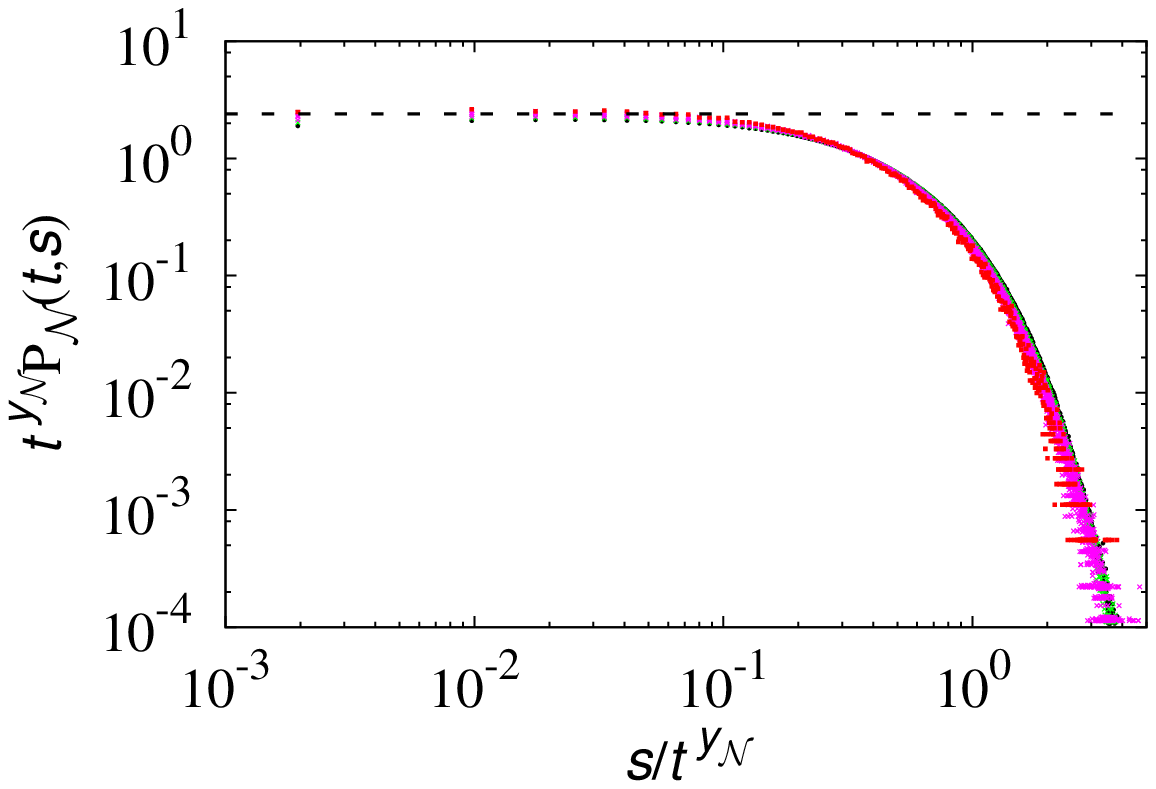}}
\subfigure[ ]{\includegraphics[scale=0.42]{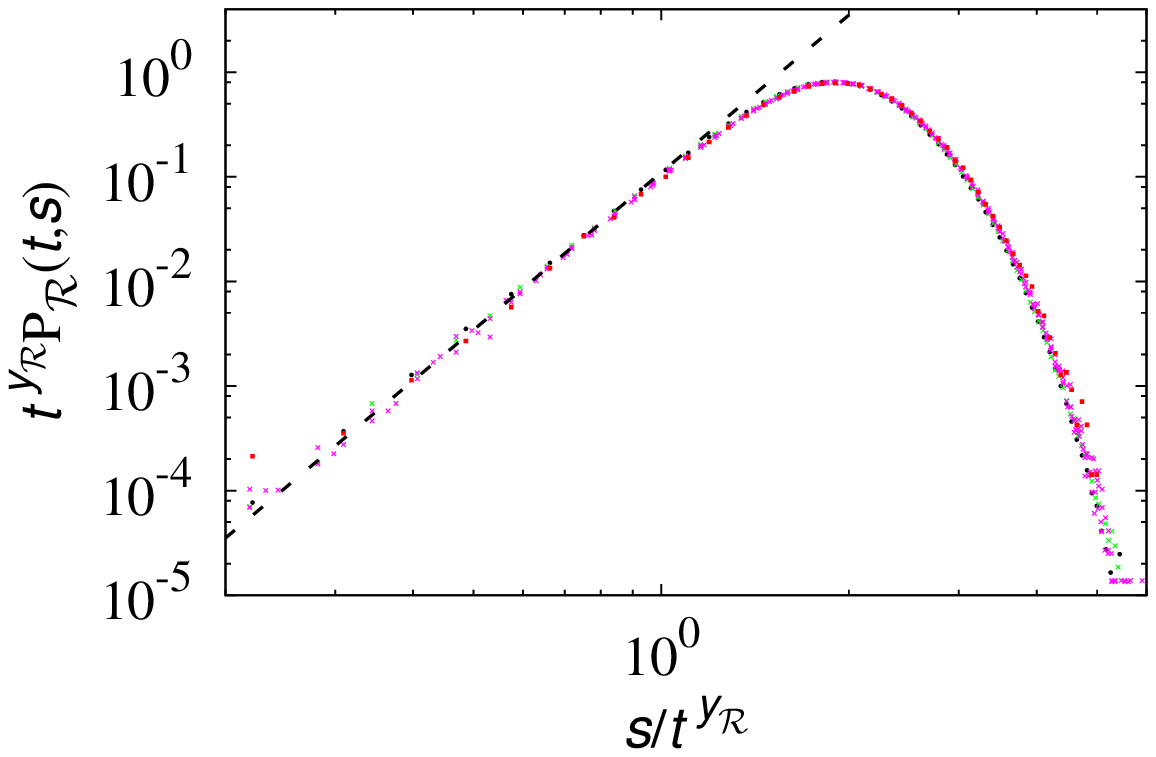}}
\subfigure[ ]{\includegraphics[scale=0.42]{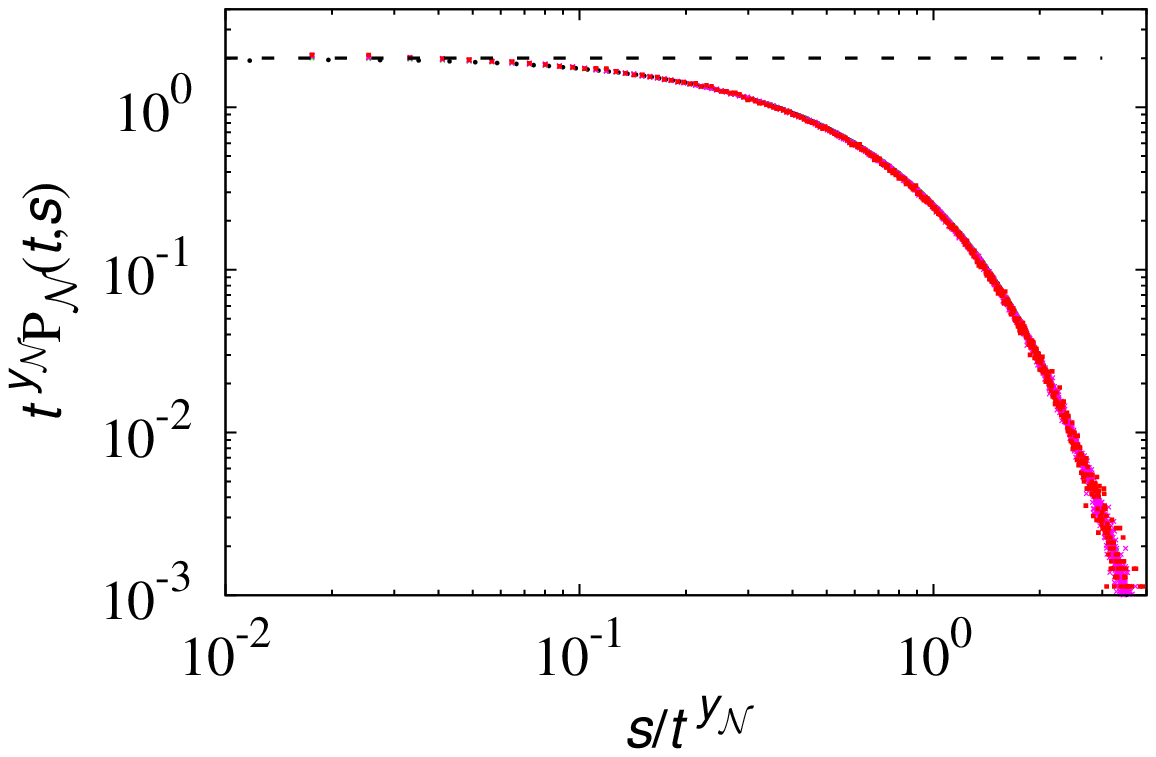}}
\subfigure[ ]{\includegraphics[scale=0.42]{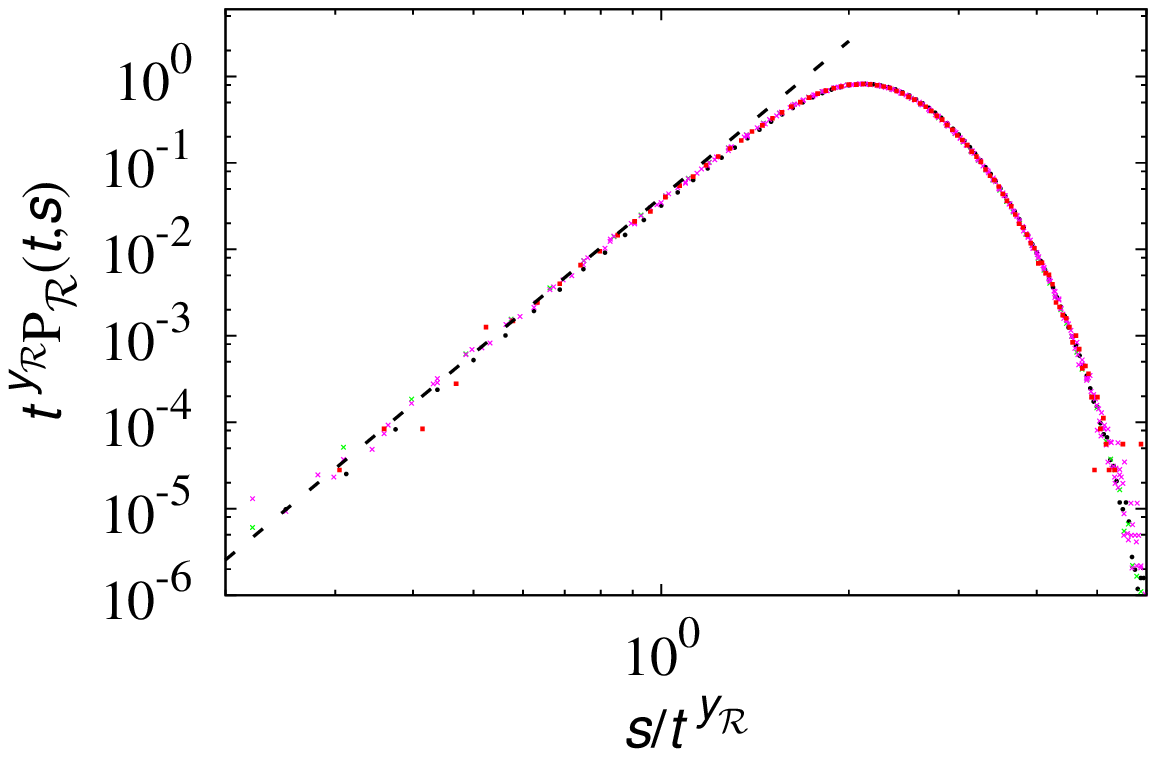}}
\caption{(Color online) Log-log plots of $t^{y_{\sN}}p_{\sN}(t,s)$ versus $s/t^{y_{\sN}}$, and $t^{y_{\sR}}p_{\sR}(t,s)$ versus $s/t^{y_{\sR}}$.
The subfigures (a) to (f) respectively correspond to $p_{\sN}(t,s)$ for $d=1$, $p_{\sR}(t,s)$ for $d=1$, $p_{\sN}(t,s)$ for $d=4$, $p_{\sR}(t,s)$ for $d=4$, $p_{\sN}(t,s)$ for $d=5$ and $p_{\sR}(t,s)$ for $d=5$.
 The data correspond to bond DP on the square lattice ($d=1$) and BCC lattice ($d=4,5$).
 The exponents $y_{\sN}=\theta+\delta$ and $y_R=1/z$ are calculated from Table~\ref{tablestandardexponents} for $d=1$, and are given by the exact mean-field values for $d=4$ and $5$.
 The dashed lines have slopes equal to $1/y_{\sN}-1$ and $1/y_{\sR} -1 + d$ for $p_{\sN}(t,s)$ and $p_{\sR}(t,s)$, respectively.}
\label{figprob}
\end{figure*}

\begin{table*}[htbp]
\begin{tabular}[t]{l|l|lllllll}
\hline
$d$        & Ref.                                     &$\beta$           &$\nupar$      & $\nu_{\perp}$    & $z$              &$\theta$      & $\delta$\\                       
\hline
{\multirow{4}{*}{$2$}}          & Present                                    &0.580(4)          &1.287(2)      & 0.729(1)         &1.7665(2)         &0.2307(2)     &0.4510(4)\\                       
           &~\cite{GrassbergerZhang1996}                &                  &1.295(6)      &                  &1.765(3)          &0.229(3)      &0.451(3) \\
           &~\cite{VoigtZiff1997}                       &                  &              &                  &1.766(2)          &0.229\,5(10)  &0.450\,5(10) \\
           &~\cite{PerlsmanHavlin2002}                  &                  &              &                  &1.766\,6(10)      &0.230\,3(4)   &0.450\,9(5)\\
\hline
{\multirow{3}{*}{$3$}}          & Present                                    & 0.818(4)         &1.106(3)      &0.582(2)          &1.8990(4)         &0.105\,7(3)   &0.739\,8(10)\\
           &~\cite{Jensen1992}                          & 0.813(11)        &1.11(1)       &                  &1.901(5)          &0.114(4)      &0.732(4) \\
           &~\cite{Janssen1981}                         & 0.822\,05        &1.105\,71     &0.583\,60         &1.887\,46         &0.120\,84     &0.737\,17\\
\hline
\end{tabular}
\caption{Final estimates of the critical exponents for $d=2$ and 3. }
\label{tablestandardexponentsD23}
\end{table*}

\section{Critical Distributions}
\label{Critical Distributions}
In this section we consider the critical scaling of $p_{\sN}(t,s)$ and $p_{\sR}(t,s)$.
From finite-size scaling theory, we expect that $p_{\sN}(t,s)$ and $p_{\sR}(t,s)$ should scale at criticality as
\begin{equation}
\label{eq:probdis}
  \begin{split}
 p_{\sN}(t,s)  &\sim t^{-y_{\sN}} F_{\sN}(s/t^{y_{\sN}}),\\
 p_{\sR}(t,s)  &\sim t^{-y_{\sR}} F_{\sR}(s/t^{y_{\sR}}).
 \end{split}
\end{equation}
 The scaling functions $F_{\sN}$ and $F_{\sR}$ are expected to be universal.
 It follows immediately from~\eqref{eq:probdis} that for all $k\in\bbN$ we have
\begin{equation}
\label{eq:scaling_NkRk} 
\begin{split}
  \<\sN_t^k\> &\sim t^{k\,y_{\sN} - \delta},\\
  \<\sR_t^k\> &\sim t^{k\,y_{\sR} - \delta}.
\end{split}
\end{equation}
Since $\<\sN_t\>\sim t^{\theta}$, we can then identify 
\begin{equation}
  y_{\sN}=\theta+\delta.
  \label{N distribution exponent}
\end{equation}
Similarly, making the assumption that 
%$$
%\left\<\frac{\sS_t^2}{\sN_t}\right\> \sim \frac{\<\sS_t^2\>}{\<\sN_t\>}
%$$
$$
\<\sR_t^2\>\sim R_t^2 \, P_t\sim t^{2/z-\delta}
$$
at criticality implies
%$\<\sR_t^2\>\sim R_t^2 \, P_t\sim t^{2/z-\delta}$, so
\begin{equation}
  y_{\sR}=1/z.
  \label{R distribution exponent}
\end{equation}

 To test these predictions, Fig.~\ref{figprob} shows log-log plots of $t^{y_{\sN}}\,p_{\sN}(t,s)$ versus $s/t^{y_{\sN}}$ and $t^{y_{\sR}}\,p_{\sR}(t,s)$ versus $s/t^{y_{\sR}}$.
 The figures show bond DP data for the square lattice for $d=1$ and the BCC lattice for $d=4,5$.
 For $d=1$, we set the exponents $y_{\sN}$ and $y_{\sR}$ to $y_{\sN}=0.473\,14$ and $y_R=0.632\,63$, using the results from Table~\ref{tablestandardexponents} in Appendix~\ref{d=1 results}.
 For $d=4$ and $5$, the mean-field predictions $y_{\sN}=1$ and $y_R=1/2$ were used.
 In principle, logarithmic corrections should be taken into account for $d=4$, however we did not pursue this here.
 The conjectures~\eqref{eq:probdis},~\eqref{N distribution exponent} and~\eqref{R distribution exponent} are strongly supported by the excellent data collapse observed in Fig.~\ref{figprob}.

 From Fig.~\ref{figprob}, we observe that for $s/t^{y_{\sN}}, s/t^{y_{\sR}} \ll 1$, the curves appear to asymptote to a straight line.
 We find empirically that these slopes are well described by the expressions $1/y_{\sN}-1$ and $1/y_R-1+d$, for $p_{\sN}(t,s)$ and $p_{\sR}(t,s)$ respectively.
 We therefore conjecture that these expressions hold exactly, and we illustrate them with the dashed lines in Fig.~\ref{figprob}.
 As a result, the scaling forms \eqref{eq:probdis} can be recast as
 \begin{equation}
   \begin{split}
  p_{\sN}(t,s)   & \sim  t^{-1}              \,s^{1/y_{\sN}-1}     \,f_{\sN}(s/t^{y_{\sN}}),\\
  p_{\sR}(t,s)   & \sim  t^{-1 - d\, y_{\sR}}\,s^{1/y_{\sR} -1 + d}\,f_{\sR}(s/t^{y_{\sR}}),
  \end{split}
  \label{probdis_rewrite}
\end{equation}
with $f_{\sN}$ and $f_{\sR}$ universal.

\section{Discussion}
\label{sec6}
 We present a high-precision Monte Carlo study of bond and site DP on $(d+1)$-dimensional simple-cubic and body-centered-cubic lattices, with $2 \leq d \leq 7$.
 A dimensionless ratio $Q_t= N_{2t}/N_t$ constructed from the number of wet sites $N_t$ is defined and used to estimate the critical thresholds.
 We report improved estimates of thresholds for $2 \le d \le 7$, and
in high dimensions ($d>4$) we provide estimates of $p_{\rm c}$ in
several cases for which no previous estimates appear to be known.
 In addition, we report improved estimates of the critical exponents for $d=2$ and $3$.
 The accuracy of these estimates was due in part to the use of reduced-variance estimators 
 introduced in~\cite{Grassberger2003,Grassberger2009b,FosterGrassbergerPaczuski2009}.
 At the estimated thresholds, we also conjecture, and numerically confirm, the finite-size 
 scaling of the critical probability distributions $p_{\sN}(t,s)$ and $p_{\sR}(t,s)$.

 The high-precision Monte Carlo data reported in this work also suggests that further investigation of a number of questions is desirable.
 Firstly, is there an underlying physical reason (e.g. hidden symmetry) that in two and three dimensions bond DP on the BCC lattice suffers less finite-size corrections than 
 site DP on the BCC lattice and both site and bond DP on the SC lattice? 
 Second, can we obtain deeper understanding of origin of the scaling behavior described by (\ref{probdis_rewrite})?

\section{Acknowledgments}
 We thank Peter Grassberger for helpful comments and for sharing code with us.
 J.F.W acknowledges the useful discussion with Wei Zhang.
 The simulations were carried out in part on NYU's ITS cluster, which is partly supported by NSF Grant No. PHY-0424082.
 In addition, this research was undertaken with the assistance of resources provided at 
 the NCI National Facility through the National Computational Merit Allocation Scheme supported by the Australian Government. 
 This work is supported by the National Nature Science Foundation of China under 
 Grant No. 91024026 and 11275185, and the Chinese Academy of Sciences.
 It was also supported under the Australian Research Council's Discovery Projects funding scheme (project number DP110101141),
 and T.G. is the recipient of an Australian Research Council Future Fellowship (project number FT100100494).
 J.F.W and Y.J.D also acknowledge the Specialized Research Fund for the Doctoral Program of Higher Education under Grant No. 20103402110053.

\appendix
\section{Estimates of thresholds and critical exponents in (1+1) dimensions.}
\label{d=1 results}
\begin{figure}
\centering
\includegraphics[scale=0.30]{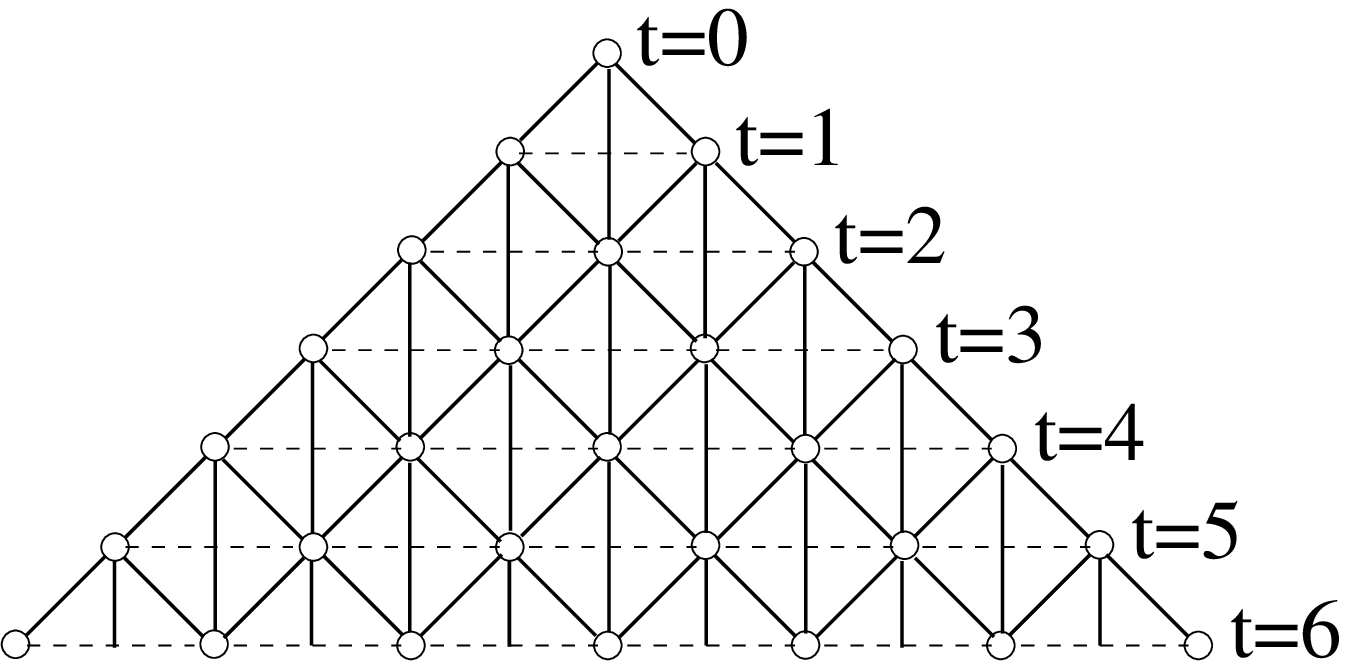}
\includegraphics[scale=0.30]{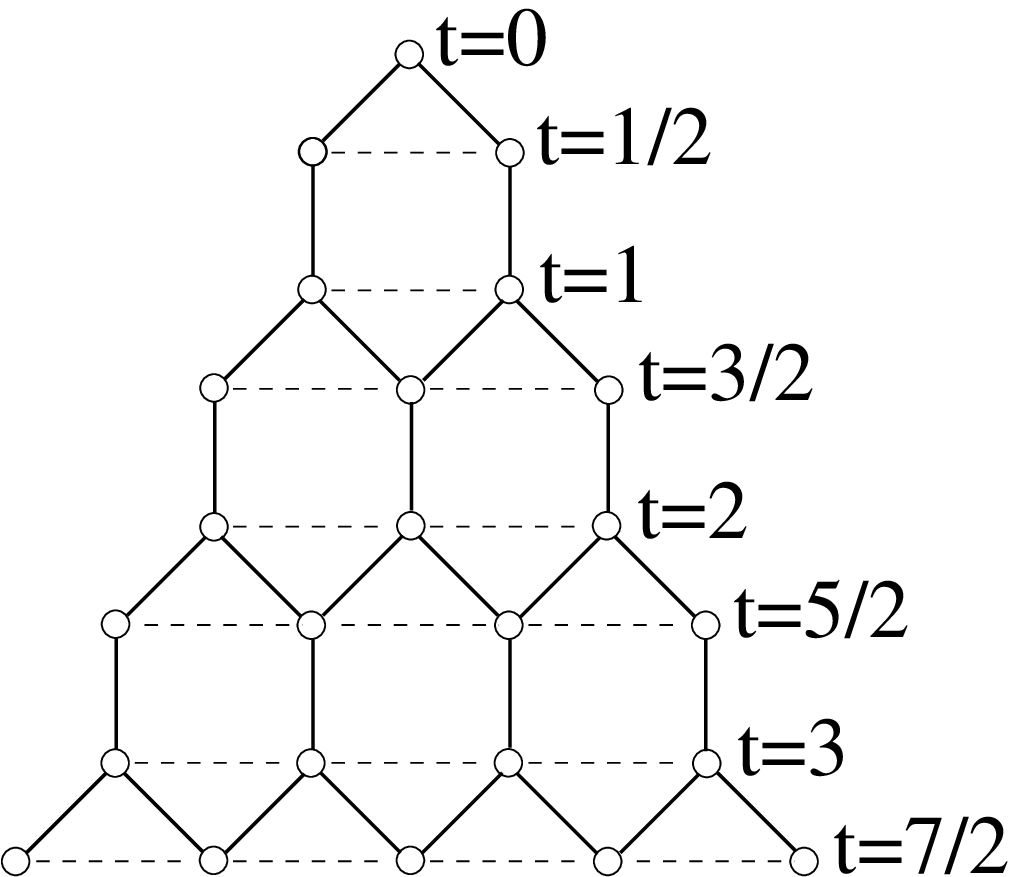}
\includegraphics[scale=0.30]{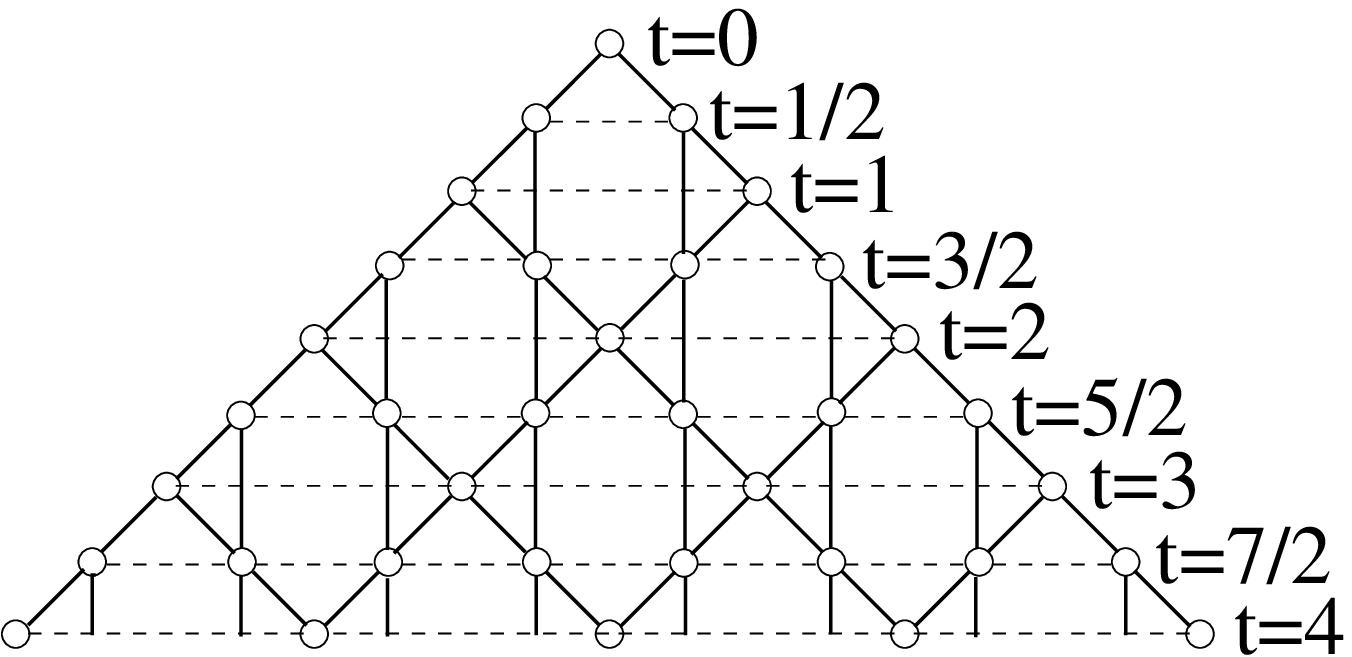}
\caption{Plots of triangular, honeycomb, and kagome lattices.}
\label{TriHexKagDiagram}
\end{figure}

In this appendix we report estimates of the critical thresholds and critical exponents for a number of $(1+1)$-dimensional lattices.
Specifically, we simulated bond and site DP on square (Fig.~\ref{squareLatticeDiagram}), triangular, honeycomb, and kagome lattices (Fig.~\ref{TriHexKagDiagram}).
On the triangular lattice, a site at time $t$ has three neighboring sites at times $t'<t$: two at $t-1$ and one at $t-2$.
On the honeycomb lattice, a site at an odd time $t$ has two neighboring sites at time $t-1$, while sites at even times have only one neighbor at time $t-1$.
On the kagome lattice, a site at an odd time $t$ has one neighbour at time $t-1$ and one at time $t-2$, while sites at even times have two neighbours at time $t-1$.

The general methodology applied for these simulations is as described in Section~\ref{Description of Simulations}.
However we did not apply the reduced-variance estimators in this case, since their variance only becomes suppressed in high dimensions.
The thresholds estimated from $Q_t$ for $d=1$ are shown in Table~\ref{tablefinalpc1d}.
The estimates of the critical exponents are shown in Table~\ref{tablestandardexponents}.
These estimate are consistent with, but less precise than, results obtained previously using series analysis.

\begin{table*}[htbp]
\begin{tabular}[t]{l|ll|lllll}
\hline
Lattice            & Site               &                                     &   & Bond               &                   \\
                   & $\pc$(Present)     & $\pc$(Previous)                     &   & $\pc$(Present)     & $\pc$(Previous)       \\
\hline
square     & 0.705\,485\,2(3)   & 0.705\,485\,22(4)~\cite{Jensen1999}  &   & 0.644\,700\,1(2)    & 0.644\,700\,185(5)~\cite{Jensen1999} \\
                  &                    & 0.705\,489(4)~\cite{LubeckWillmann2002}      &   &                    & 0.644\,700\,15(5)~\cite{Jensen1996}  \\
\hline
triangular & 0.595\,647\,0(3)   & 0.595\,646\,75(10)~\cite{Jensen2004} &   & 0.478\,025\,0(4)  & 0.478\,025\,25(5)~\cite{Jensen2004} \\
                  &             & 0.595\,646\,8(5)~\cite{Jensen1996}      &   &                    & 0.478\,025(1)~\cite{Jensen1996}  \\
\hline
honeycomb  & 0.839\,931\,6(2)   & 0.839\,933(5)~\cite{JensenGuttmann1995}  &   & 0.822\,856\,9(2)      & 0.822\,856\,80(6)~\cite{Jensen2004} \\
\hline
kagome   & 0.736\,931\,7(2)   & 0.736\,931\,82(4)~\cite{Jensen2004}  &   & 0.658\,968\,9(2)  & 0.658\,969\,10(8)~\cite{Jensen2004} \\
\hline
\end{tabular}
\caption{Estimates of thresholds in (1+1) dimensions on the square, triangular, honeycomb and kagome lattices.}
\label{tablefinalpc1d}
\end{table*}

\begin{table*}[htbp]
\begin{tabular}[t]{llllllll}
\hline
                          &$\beta$           &$\nupar$           & $\nu_{\perp}$    & $z$              &$\theta$      & $\delta$\\
Present                   & 0.276\,7(3)        &1.735\,5(15)         &1.097\,9(10)         &1.580\,7(2)        &0.313\,70(5)    & 0.159\,44(2)\\
~\cite{Jensen1999}       & 0.276\,486(8)      &1.733\,847(6)        &1.096\,854(4)       &1.580\,745(10)      &0.313\,686(8)   &0.159\,464(6)\\ 
\hline
\end{tabular}
\caption{Estimates of the critical exponents for $d=1$.}
\label{tablestandardexponents}
\end{table*}

\section{Discussion of the improved estimators}
\label{proofs relating to estimators}
In this appendix we prove the identities~\eqref{N alternative estimator identity} and~\eqref{R alternative estimator identity}. Both are direct consequences of the following lemma.
\begin{lemma}
For both bond and site DP we have the following.
If $b_v$ is the number of Bernoulli trials required to determine the state of $v\in V_t$ given the site configuration at time $t-1$, then
$$
\bbP(s_v=1) = p\,\< b_v \>.
$$
\label{Bernoulli mean lemma}
\end{lemma}

It follows immediately from Lemma~\ref{Bernoulli mean lemma} that for any set of constants $a_v$ with $v\in V_t$ we have
\begin{equation}
\left\<\sum_{v\in V_t}a_v\,\delta_{s_v,1}\right\> = p \left\<\sum_{v\in V_t}a_v\,b_v\right\>,
\label{summed Bernoulli mean identity}
\end{equation}
where $\delta_{\cdot,\cdot}$ denotes the Kronecker delta. Choosing $a_v=1$ in~\eqref{summed Bernoulli mean identity} gives~\eqref{N alternative estimator identity}, while choosing
$a_v=r_v^2$ gives~\eqref{R alternative estimator identity}.

It now remains only to prove Lemma~\ref{Bernoulli mean lemma}.
\begin{proof}[Proof of Lemma~\ref{Bernoulli mean lemma}]
  For $v\in V_{t}$, let $n_v$ denote the number of wet neighbours of $v$ in $V_{t-1}$.

  For site DP, 
  $$
  b_v = 
  \begin{cases}
    1, & \text{ if } n_v>0,\\
    0, & \text{ if } n_v = 0,
    \end{cases}
  $$
  and so $\< b_v\> = \bbP(n_v>0)$. Since $\bbP(s_v=1)=p\,\bbP(n_v>0)$, the stated result then follows.

 For bond DP, the situation is more involved.
 Since $\bbP(s_v =1) =\left\< 1-(1-p)^{n_v}\right\>$, our task is to establish
 \begin{equation}
   p\left\< b_v \right\>
   = 
   \left\< 1-(1-p)^{n_v}\right\>.
   \label{bond identity for Bernoulli mean proof}
 \end{equation}
 If we consider a fixed value of $n_v$ then consideration of the stochastic process defined in Section~\ref{Generating DP configurations} shows that
 \begin{eqnarray}
 \bbE(b_v|n_v) &=& \sum_{k=1}^{n_v - 1}k(1-p)^{k-1}p + n_v(1-p)^{n_v - 1}\nonumber \\
 &=& \dfrac{1}{p}(1 - (1-p)^{n_v})\;.
 \label{bond conditional expectation identity}
 \end{eqnarray}
 From~\eqref{bond conditional expectation identity} 
 Taking the expectation of~\eqref{bond conditional expectation identity} yields~\eqref{bond identity for Bernoulli mean proof}, which concludes the proof.
\end{proof}

%

%\bibliography{bib.bib}

\begin{thebibliography}{31}%
\makeatletter
\providecommand \@ifxundefined [1]{%
 \@ifx{#1\undefined}
}%
\providecommand \@ifnum [1]{%
 \ifnum #1\expandafter \@firstoftwo
 \else \expandafter \@secondoftwo
 \fi
}%
\providecommand \@ifx [1]{%
 \ifx #1\expandafter \@firstoftwo
 \else \expandafter \@secondoftwo
 \fi
}%
\providecommand \natexlab [1]{#1}%
\providecommand \enquote  [1]{``#1''}%
\providecommand \bibnamefont  [1]{#1}%
\providecommand \bibfnamefont [1]{#1}%
\providecommand \citenamefont [1]{#1}%
\providecommand \href@noop [0]{\@secondoftwo}%
\providecommand \href [0]{\begingroup \@sanitize@url \@href}%
\providecommand \@href[1]{\@@startlink{#1}\@@href}%
\providecommand \@@href[1]{\endgroup#1\@@endlink}%
\providecommand \@sanitize@url [0]{\catcode `\\12\catcode `\$12\catcode
  `\&12\catcode `\#12\catcode `\^12\catcode `\_12\catcode `\%12\relax}%
\providecommand \@@startlink[1]{}%
\providecommand \@@endlink[0]{}%
\providecommand \url  [0]{\begingroup\@sanitize@url \@url }%
\providecommand \@url [1]{\endgroup\@href {#1}{\urlprefix }}%
\providecommand \urlprefix  [0]{URL }%
\providecommand \Eprint [0]{\href }%
\providecommand \doibase [0]{http://dx.doi.org/}%
\providecommand \selectlanguage [0]{\@gobble}%
\providecommand \bibinfo  [0]{\@secondoftwo}%
\providecommand \bibfield  [0]{\@secondoftwo}%
\providecommand \translation [1]{[#1]}%
\providecommand \BibitemOpen [0]{}%
\providecommand \bibitemStop [0]{}%
\providecommand \bibitemNoStop [0]{.\EOS\space}%
\providecommand \EOS [0]{\spacefactor3000\relax}%
\providecommand \BibitemShut  [1]{\csname bibitem#1\endcsname}%
\let\auto@bib@innerbib\@empty
\bibitem [{\citenamefont {Broadbent}\ and\ \citenamefont
  {Hammersley}(1957)}]{BroadbentHarmmersley1957}%
  \BibitemOpen
  \bibfield  {author} {\bibinfo {author} {\bibfnamefont {S.~R.}\ \bibnamefont
  {Broadbent}}\ and\ \bibinfo {author} {\bibfnamefont {J.~M.}\ \bibnamefont
  {Hammersley}},\ }\href@noop {} {\bibfield  {journal} {\bibinfo  {journal}
  {Proceedings of the Cambridge Philosophical Society}\ }\textbf {\bibinfo
  {volume} {53}},\ \bibinfo {pages} {629} (\bibinfo {year} {1957})}\BibitemShut
  {NoStop}%
\bibitem [{\citenamefont {Albano}(1994)}]{Albano1994}%
  \BibitemOpen
  \bibfield  {author} {\bibinfo {author} {\bibfnamefont {E.~V.}\ \bibnamefont
  {Albano}},\ }\href@noop {} {\bibfield  {journal} {\bibinfo  {journal} {J.
  Phys. A}\ }\textbf {\bibinfo {volume} {27}},\ \bibinfo {pages} {L881}
  (\bibinfo {year} {1994})}\BibitemShut {NoStop}%
\bibitem [{\citenamefont {Mollison}(1977)}]{Mollison1977}%
  \BibitemOpen
  \bibfield  {author} {\bibinfo {author} {\bibfnamefont {D.}~\bibnamefont
  {Mollison}},\ }\href@noop {} {\bibfield  {journal} {\bibinfo  {journal} {J.
  R. Stat. Soc. Ser. B (Methodol)}\ }\textbf {\bibinfo {volume} {39}},\
  \bibinfo {pages} {283} (\bibinfo {year} {1977})}\BibitemShut {NoStop}%
\bibitem [{\citenamefont {Bouchaud}\ and\ \citenamefont
  {Georges}(1990)}]{BouchaudGeorges1990}%
  \BibitemOpen
  \bibfield  {author} {\bibinfo {author} {\bibfnamefont {J.-P.}\ \bibnamefont
  {Bouchaud}}\ and\ \bibinfo {author} {\bibfnamefont {A.}~\bibnamefont
  {Georges}},\ }\href@noop {} {\bibfield  {journal} {\bibinfo  {journal} {Phys.
  Rep.}\ }\textbf {\bibinfo {volume} {195}},\ \bibinfo {pages} {127} (\bibinfo
  {year} {1990})}\BibitemShut {NoStop}%
\bibitem [{\citenamefont {Havlin}\ and\ \citenamefont
  {Benavraham}(1987)}]{HavlinBenavraham1987}%
  \BibitemOpen
  \bibfield  {author} {\bibinfo {author} {\bibfnamefont {S.}~\bibnamefont
  {Havlin}}\ and\ \bibinfo {author} {\bibfnamefont {D.}~\bibnamefont
  {Benavraham}},\ }\href@noop {} {\bibfield  {journal} {\bibinfo  {journal}
  {Adv. Phys.}\ }\textbf {\bibinfo {volume} {36}},\ \bibinfo {pages} {695}
  (\bibinfo {year} {1987})}\BibitemShut {NoStop}%
\bibitem [{\citenamefont {Janssen}(1981)}]{Janssen1981}%
  \BibitemOpen
  \bibfield  {author} {\bibinfo {author} {\bibfnamefont {H.-K.}\ \bibnamefont
  {Janssen}},\ }\href@noop {} {\bibfield  {journal} {\bibinfo  {journal} {Z.
  Phys. B}\ }\textbf {\bibinfo {volume} {42}},\ \bibinfo {pages} {151}
  (\bibinfo {year} {1981})}\BibitemShut {NoStop}%
\bibitem [{\citenamefont {Grassberger}(1982)}]{Grassberger1982}%
  \BibitemOpen
  \bibfield  {author} {\bibinfo {author} {\bibfnamefont {P.}~\bibnamefont
  {Grassberger}},\ }\href@noop {} {\bibfield  {journal} {\bibinfo  {journal}
  {Z. Phys. B}\ }\textbf {\bibinfo {volume} {47}},\ \bibinfo {pages} {365}
  (\bibinfo {year} {1982})}\BibitemShut {NoStop}%
\bibitem [{\citenamefont {Jensen}(1996)}]{Jensen1996}%
  \BibitemOpen
  \bibfield  {author} {\bibinfo {author} {\bibfnamefont {I.}~\bibnamefont
  {Jensen}},\ }\href@noop {} {\bibfield  {journal} {\bibinfo  {journal} {J.
  Phys. A}\ }\textbf {\bibinfo {volume} {29}},\ \bibinfo {pages} {7013}
  (\bibinfo {year} {1996})}\BibitemShut {NoStop}%
\bibitem [{\citenamefont {Jensen}(1999)}]{Jensen1999}%
  \BibitemOpen
  \bibfield  {author} {\bibinfo {author} {\bibfnamefont {I.}~\bibnamefont
  {Jensen}},\ }\href@noop {} {\bibfield  {journal} {\bibinfo  {journal} {J.
  Phys. A}\ }\textbf {\bibinfo {volume} {32}},\ \bibinfo {pages} {5233}
  (\bibinfo {year} {1999})}\BibitemShut {NoStop}%
\bibitem [{\citenamefont {Grassberger}\ and\ \citenamefont
  {Zhang}(1996)}]{GrassbergerZhang1996}%
  \BibitemOpen
  \bibfield  {author} {\bibinfo {author} {\bibfnamefont {P.}~\bibnamefont
  {Grassberger}}\ and\ \bibinfo {author} {\bibfnamefont {Y.}~\bibnamefont
  {Zhang}},\ }\href@noop {} {\bibfield  {journal} {\bibinfo  {journal} {physica
  A}\ }\textbf {\bibinfo {volume} {224}},\ \bibinfo {pages} {169} (\bibinfo
  {year} {1996})}\BibitemShut {NoStop}%
\bibitem [{\citenamefont {Grassberger}(2009{\natexlab{a}})}]{Grassberger2009a}%
  \BibitemOpen
  \bibfield  {author} {\bibinfo {author} {\bibfnamefont {P.}~\bibnamefont
  {Grassberger}},\ }\href@noop {} {\bibfield  {journal} {\bibinfo  {journal}
  {J. Stat. Mech:. Theory Exp.}\ ,\ \bibinfo {pages} {P08021}} (\bibinfo {year}
  {2009}{\natexlab{a}})}\BibitemShut {NoStop}%
\bibitem [{\citenamefont {Perlsman}\ and\ \citenamefont
  {Havlin}(2002)}]{PerlsmanHavlin2002}%
  \BibitemOpen
  \bibfield  {author} {\bibinfo {author} {\bibfnamefont {E.}~\bibnamefont
  {Perlsman}}\ and\ \bibinfo {author} {\bibfnamefont {S.}~\bibnamefont
  {Havlin}},\ }\href@noop {} {\bibfield  {journal} {\bibinfo  {journal}
  {Europhys. Lett.}\ }\textbf {\bibinfo {volume} {58}},\ \bibinfo {pages} {176}
  (\bibinfo {year} {2002})}\BibitemShut {NoStop}%
\bibitem [{\citenamefont {Lubeck}\ and\ \citenamefont
  {Willmann}(2004)}]{LubeckWillmann2004}%
  \BibitemOpen
  \bibfield  {author} {\bibinfo {author} {\bibfnamefont {S.}~\bibnamefont
  {Lubeck}}\ and\ \bibinfo {author} {\bibfnamefont {R.}~\bibnamefont
  {Willmann}},\ }\href@noop {} {\bibfield  {journal} {\bibinfo  {journal} {J.
  Stat. Phys.}\ }\textbf {\bibinfo {volume} {115}},\ \bibinfo {pages} {1231}
  (\bibinfo {year} {2004})}\BibitemShut {NoStop}%
\bibitem [{\citenamefont {Adler}\ \emph {et~al.}(1988)\citenamefont {Adler},
  \citenamefont {Berger}, \citenamefont {Duarte},\ and\ \citenamefont
  {Meir}}]{AdlerBergerDuarteMeir1988}%
  \BibitemOpen
  \bibfield  {author} {\bibinfo {author} {\bibfnamefont {J.}~\bibnamefont
  {Adler}}, \bibinfo {author} {\bibfnamefont {J.}~\bibnamefont {Berger}},
  \bibinfo {author} {\bibfnamefont {J.~A. M.~S.}\ \bibnamefont {Duarte}}, \
  and\ \bibinfo {author} {\bibfnamefont {Y.}~\bibnamefont {Meir}},\ }\href@noop
  {} {\bibfield  {journal} {\bibinfo  {journal} {Phys. Rev. B}\ }\textbf
  {\bibinfo {volume} {37}},\ \bibinfo {pages} {7529} (\bibinfo {year}
  {1988})}\BibitemShut {NoStop}%
\bibitem [{\citenamefont {Blease}(1977)}]{Blease1977}%
  \BibitemOpen
  \bibfield  {author} {\bibinfo {author} {\bibfnamefont {J.}~\bibnamefont
  {Blease}},\ }\href@noop {} {\bibfield  {journal} {\bibinfo  {journal} {J.
  Phys. C}\ }\textbf {\bibinfo {volume} {10}},\ \bibinfo {pages} {917}
  (\bibinfo {year} {1977})}\BibitemShut {NoStop}%
\bibitem [{\citenamefont {Grassberger}(2009{\natexlab{b}})}]{Grassberger2009b}%
  \BibitemOpen
  \bibfield  {author} {\bibinfo {author} {\bibfnamefont {P.}~\bibnamefont
  {Grassberger}},\ }\href@noop {} {\bibfield  {journal} {\bibinfo  {journal}
  {Phys. Rev. E}\ }\textbf {\bibinfo {volume} {79}},\ \bibinfo {pages} {052104}
  (\bibinfo {year} {2009}{\natexlab{b}})}\BibitemShut {NoStop}%
\bibitem [{Note1()}]{Note1}%
  \BibitemOpen
  \bibinfo {note} {We note that the version of bond DP that we are simulating
  generates a different ensemble of bond configurations compared to the
  standard geometric version of bond DP, in which each edge is occupied
  independently. However, the resulting site configurations generated by these
  two bond DP models are identical. Since we only consider properties of the
  site configurations in this article, the distinction is unimportant for our
  purposes. For the sake of computational efficiency, we find the version
  described in the text more convenient.}\BibitemShut {Stop}%
\bibitem [{\citenamefont {Sedgewick}(1998)}]{Sedgewick98}%
  \BibitemOpen
  \bibfield  {author} {\bibinfo {author} {\bibfnamefont {R.}~\bibnamefont
  {Sedgewick}},\ }\href@noop {} {\emph {\bibinfo {title} {Algorithms in C}}},\
  \bibinfo {edition} {3rd}\ ed.\ (\bibinfo  {publisher} {Addison-Wesley},\
  \bibinfo {address} {Reading, Massachusetts},\ \bibinfo {year}
  {1998})\BibitemShut {NoStop}%
\bibitem [{\citenamefont {Grassberger}(2003)}]{Grassberger2003}%
  \BibitemOpen
  \bibfield  {author} {\bibinfo {author} {\bibfnamefont {P.}~\bibnamefont
  {Grassberger}},\ }\href@noop {} {\bibfield  {journal} {\bibinfo  {journal}
  {Phys. Rev. E}\ }\textbf {\bibinfo {volume} {67}},\ \bibinfo {pages} {036101}
  (\bibinfo {year} {2003})}\BibitemShut {NoStop}%
\bibitem [{\citenamefont {Foster}\ \emph {et~al.}(2009)\citenamefont {Foster},
  \citenamefont {Grassberger},\ and\ \citenamefont
  {Paczuski}}]{FosterGrassbergerPaczuski2009}%
  \BibitemOpen
  \bibfield  {author} {\bibinfo {author} {\bibfnamefont {J.~G.}\ \bibnamefont
  {Foster}}, \bibinfo {author} {\bibfnamefont {P.}~\bibnamefont {Grassberger}},
  \ and\ \bibinfo {author} {\bibfnamefont {M.}~\bibnamefont {Paczuski}},\
  }\href@noop {} {\bibfield  {journal} {\bibinfo  {journal} {New J. Phys.}\
  }\textbf {\bibinfo {volume} {{11}}},\ \bibinfo {pages} {023009} (\bibinfo
  {year} {{2009}})}\BibitemShut {NoStop}%
\bibitem [{\citenamefont {Dickman}(1999)}]{Dickman1999}%
  \BibitemOpen
  \bibfield  {author} {\bibinfo {author} {\bibfnamefont {R.}~\bibnamefont
  {Dickman}},\ }\href@noop {} {\bibfield  {journal} {\bibinfo  {journal} {Phys.
  Rev. E}\ }\textbf {\bibinfo {volume} {60}},\ \bibinfo {pages} {R2441}
  (\bibinfo {year} {1999})}\BibitemShut {NoStop}%
\bibitem [{\citenamefont {Janssen}\ and\ \citenamefont
  {T\"auber}(2005)}]{JanssenTauber2005}%
  \BibitemOpen
  \bibfield  {author} {\bibinfo {author} {\bibfnamefont {H.-K.}\ \bibnamefont
  {Janssen}}\ and\ \bibinfo {author} {\bibfnamefont {U.}~\bibnamefont
  {T\"auber}},\ }\href@noop {} {\bibfield  {journal} {\bibinfo  {journal} {Ann.
  Phys.}\ }\textbf {\bibinfo {volume} {315}},\ \bibinfo {pages} {147} (\bibinfo
  {year} {2005})}\BibitemShut {NoStop}%
\bibitem [{\citenamefont {Janssen}\ and\ \citenamefont
  {Stenull}(2004)}]{JanssenStenull2004}%
  \BibitemOpen
  \bibfield  {author} {\bibinfo {author} {\bibfnamefont {H.-K.}\ \bibnamefont
  {Janssen}}\ and\ \bibinfo {author} {\bibfnamefont {O.}~\bibnamefont
  {Stenull}},\ }\href@noop {} {\bibfield  {journal} {\bibinfo  {journal} {Phys.
  Rev. E}\ }\textbf {\bibinfo {volume} {69}},\ \bibinfo {pages} {016125}
  (\bibinfo {year} {2004})}\BibitemShut {NoStop}%
\bibitem [{\citenamefont {Hinrichsen}(2000)}]{Hinrichsen2000a}%
  \BibitemOpen
  \bibfield  {author} {\bibinfo {author} {\bibfnamefont {H.}~\bibnamefont
  {Hinrichsen}},\ }\href@noop {} {\bibfield  {journal} {\bibinfo  {journal}
  {Adv. Phys.}\ }\textbf {\bibinfo {volume} {49}},\ \bibinfo {pages} {815}
  (\bibinfo {year} {2000})}\BibitemShut {NoStop}%
\bibitem [{\citenamefont {Kurrer}\ and\ \citenamefont
  {Schulten}(1993)}]{KurrerSchulten1993}%
  \BibitemOpen
  \bibfield  {author} {\bibinfo {author} {\bibfnamefont {C.}~\bibnamefont
  {Kurrer}}\ and\ \bibinfo {author} {\bibfnamefont {K.}~\bibnamefont
  {Schulten}},\ }\href@noop {} {\bibfield  {journal} {\bibinfo  {journal}
  {Phys. Rev. E}\ }\textbf {\bibinfo {volume} {48}},\ \bibinfo {pages} {614}
  (\bibinfo {year} {1993})}\BibitemShut {NoStop}%
\bibitem [{\citenamefont {Bronzan}\ and\ \citenamefont
  {Dash}(1974)}]{BronzanDash1974}%
  \BibitemOpen
  \bibfield  {author} {\bibinfo {author} {\bibfnamefont {J.}~\bibnamefont
  {Bronzan}}\ and\ \bibinfo {author} {\bibfnamefont {J.}~\bibnamefont {Dash}},\
  }\href@noop {} {\bibfield  {journal} {\bibinfo  {journal} {Phys. Lett. B}\
  }\textbf {\bibinfo {volume} {B 51}},\ \bibinfo {pages} {496} (\bibinfo {year}
  {1974})}\BibitemShut {NoStop}%
\bibitem [{\citenamefont {Voigt}\ and\ \citenamefont
  {Ziff}(1997)}]{VoigtZiff1997}%
  \BibitemOpen
  \bibfield  {author} {\bibinfo {author} {\bibfnamefont {C.~A.}\ \bibnamefont
  {Voigt}}\ and\ \bibinfo {author} {\bibfnamefont {R.~M.}\ \bibnamefont
  {Ziff}},\ }\href@noop {} {\bibfield  {journal} {\bibinfo  {journal} {Phys.
  Rev. E}\ }\textbf {\bibinfo {volume} {56}},\ \bibinfo {pages} {R6241}
  (\bibinfo {year} {1997})}\BibitemShut {NoStop}%
\bibitem [{\citenamefont {Jensen}(1992)}]{Jensen1992}%
  \BibitemOpen
  \bibfield  {author} {\bibinfo {author} {\bibfnamefont {I.}~\bibnamefont
  {Jensen}},\ }\href@noop {} {\bibfield  {journal} {\bibinfo  {journal} {Phys.
  Rev. A}\ }\textbf {\bibinfo {volume} {45}},\ \bibinfo {pages} {R563}
  (\bibinfo {year} {1992})}\BibitemShut {NoStop}%
\bibitem [{\citenamefont {Lubeck}\ and\ \citenamefont
  {Willmann}(2002)}]{LubeckWillmann2002}%
  \BibitemOpen
  \bibfield  {author} {\bibinfo {author} {\bibfnamefont {S.}~\bibnamefont
  {Lubeck}}\ and\ \bibinfo {author} {\bibfnamefont {R.}~\bibnamefont
  {Willmann}},\ }\href@noop {} {\bibfield  {journal} {\bibinfo  {journal} {J.
  Phys. A}\ }\textbf {\bibinfo {volume} {35}},\ \bibinfo {pages} {10205}
  (\bibinfo {year} {2002})}\BibitemShut {NoStop}%
\bibitem [{\citenamefont {Jensen}(2004)}]{Jensen2004}%
  \BibitemOpen
  \bibfield  {author} {\bibinfo {author} {\bibfnamefont {I.}~\bibnamefont
  {Jensen}},\ }\href@noop {} {\bibfield  {journal} {\bibinfo  {journal} {J.
  Phys. A}\ }\textbf {\bibinfo {volume} {37}},\ \bibinfo {pages} {6899}
  (\bibinfo {year} {2004})}\BibitemShut {NoStop}%
\bibitem [{\citenamefont {Jensen}\ and\ \citenamefont
  {Guttmann}(1995)}]{JensenGuttmann1995}%
  \BibitemOpen
  \bibfield  {author} {\bibinfo {author} {\bibfnamefont {I.}~\bibnamefont
  {Jensen}}\ and\ \bibinfo {author} {\bibfnamefont {A.}~\bibnamefont
  {Guttmann}},\ }\href@noop {} {\bibfield  {journal} {\bibinfo  {journal} {J.
  Phys. A}\ }\textbf {\bibinfo {volume} {28}},\ \bibinfo {pages} {4813}
  (\bibinfo {year} {1995})}\BibitemShut {NoStop}%
\end{thebibliography}
\end{document}